\documentclass[useAMS,usenatbib,aas_macros]{mnras}
\usepackage{graphicx}
\usepackage{amssymb}
\usepackage{footnote}

\usepackage{epsfig}
\usepackage{color}
\usepackage{amssymb}
\usepackage{ams math}
\usepackage[greek,english]{babel}
\usepackage{subfigure}
\usepackage{relsize}
\usepackage{xfrac}
\usepackage{placeins,lscape}

\def\ciii{\mbox{C\,{\sc iii]\sc{$\lambda$1909}}}}

\def\hii{\mbox{H\,{\sc ii}}}

\def\halpha{\mbox{H\,{\sc $\alpha$}}}
\def\hbeta{\mbox{H\,{\sc $\beta$}}}

\title[Dust attenuation at $z=3-4$]{The VANDELS survey: Dust attenuation in star-forming galaxies at $\mathbf{z=3-4}$}
\author[F. Cullen et al.]{F. Cullen$^{1}$\thanks{E-mail:fc@roe.ac.uk}, 
R. J. McLure${^{1}}$, S. Khochfar${^{1}}$, J. S. Dunlop${^{1}}$, C. Dalla Vecchia${^{2,3}}$,
\and A. C. Carnall${^{1}}$, N. Bourne${^{1}}$, M. Castellano${^{4}}$, A. Cimatti${^{5,6}}$,
M. Cirasuolo${^{7}}$, D. Elbaz${^{8}}$, \and J. P. U. Fynbo${^{9}}$, B. Garilli${^{10}}$, A. Koekemoer${^{11}}$, F. Marchi${^{4}}$, L. Pentericci${^{4}}$, M. Talia${^{5,12}}$, \and G. Zamorani${^{12}}$\\
$^{1}$SUPA\thanks{Scottish Universities Physics Alliance}, Institute for Astronomy, University of Edinburgh, Royal Observatory, Edinburgh EH9 3HJ\\
$^{2}$Instituto de Astrof\'isica de Canarias, C/V\'ia L\'actea s/n, E-38205 La Laguna, Tenerife, Spain\\
$^{3}$Departamento de Astrof\'isica, Universidad de La Laguna, Av. del Astrof\'isico Franciso S\'anchex s/n, E-38206 La Laguna, Tenerife, Spain\\
$^{4}$INAF$-$Osservatorio Astronomico di Roma, Via Frascati 33, I-00040 Monte Porzio Catone (RM), Italy\\
$^{5}$Dipartimento di Fisica e Astronomia, Universit\`a de Bologna, Viale Gobetti 93/2, I-40129 Bologna, Italy\\
$^{6}$INAF $-$ Osservatorio Astrofisico di Arcetri, Largo E. Fermi 5, I-50125, Firenze, Italy \\
$^{7}$European Southern Observatory, Karl-Schwarzschild-Str. 2, 86748 Garching b. M\"unchen, Germany\\
$^{8}$Laboratoire AIM-Paris-Saclay, CEA/DSM/Irfu - CNRS - Universit\'e Paris Diderot\\
$^{9}$The Cosmic Dawn Center, Niels Bohr Institute, University of Copenhagen, Juliane Maries Vej 30, DK-2100 Copenhagen {{\O}}, Denmark\\
$^{10}$INAF-IASF Milano, via Bassini 15, I-20133, Milano, Italy\\
$^{11}$Space Telescope Science Institute, 3700 San Martin Dr., Baltimore MD 21218, USA\\
$^{12}$INAF- Osservatorio Astronomico di Bologna, Via Gobetti 93/3, I-40129, Bologna, Italy}

\begin{document}

\date{Accepted -- . Received 2017 May 30}

\pagerange{\pageref{firstpage}--\pageref{lastpage}} \pubyear{2016}

\maketitle	

\label{firstpage}

\begin{abstract}
We present the results of a new study of dust attenuation at redshifts $3 < z < 4$ based on a sample of $236$ star-forming galaxies from the VANDELS spectroscopic survey.
Motivated by results from the First Billion Years (FiBY) simulation project, we argue that the intrinsic spectral energy distributions (SEDs) of star-forming galaxies at these redshifts have a self-similar shape across the mass range $8.2 \leq$ log$(M_{\star}/M_{\odot}) \leq 10.6$ probed by our sample.
Using FiBY data, we construct a set of intrinsic SED templates which incorporate both detailed star formation and chemical abundance histories, and a variety of stellar population synthesis (SPS) model assumptions.
With this set of intrinsic SEDs, we present a novel approach for directly recovering the shape and normalization of the dust attenuation curve.
We find, across all of the intrinsic templates considered, that the average attenuation curve for star-forming galaxies at $z\simeq3.5$ is similar in shape to the commonly-adopted Calzetti starburst law, with an average total-to-selective attenuation ratio of $R_{V}=4.18\pm0.29$.
In contrast, we find that an average attenuation curve as steep as the SMC extinction law is strongly disfavoured.
We show that the optical attenuation ($A_V$) versus stellar mass ($M_{\star}$) relation predicted using our method is consistent with recent ALMA observations of galaxies at $2<z<3$ in the \emph{Hubble} \emph{Ultra} \emph{Deep} \emph{Field} (HUDF), as well as empirical $A_V - M_{\star}$ relations predicted by a Calzetti-like law.
In fact, our results, combined with other literature data, suggest that the $A_V - M_{\star}$ relation does not evolve over the redshift range $0<z<5$, at least for galaxies with log$(M_{\star}/M_{\odot}) \gtrsim 9.5$.
Finally, we present tentative evidence which suggests that the attenuation curve may become steeper at lower masses log$(M_{\star}/M_{\odot}) \lesssim 9.0$.
\end{abstract} 



\section{Introduction}

Interstellar dust absorbs and scatters the ultraviolet to near-infrared (near-IR) radiation emitted by stars in galaxies, and re-radiates the absorbed energy in the infrared (IR) and far-infrared (FIR).
The observed spectral energy distribution (SED) of a galaxy is therefore strongly affected by the amount, and spatial distribution, of the dust within it.
Understanding exactly how dust influences the SEDs of galaxies is crucial to our interpretation of observations.

Characterizing the line-of-sight extinction towards individual stars has been one method used to determine the wavelength-dependent effect of dust in the Milky Way and other nearby galaxies, referred to as the extinction curve \citep[e.g][]{prevot1984, cardelli1989, gordon2003,hagen2017}.
In the case of an extinction curve, we are sensitive to both absorption and scattering out of the line-of-sight by dust.
Uniformly across all curves, the extinction peaks in the ultraviolet (UV) and falls off with increasing wavelength, a shape which is related to the size distribution and composition of the dust grains \citep[e.g.][]{draine2003}.

For distant galaxies it is in general not possible to measure the extinction of stars along individual sight-lines. Exceptions do exist, for example in the case of gamma-ray burst afterglow spectra \citep[e.g.][]{zafar2011,fynbo2014,heintz2017}; however, in the vast majority of cases, we are observing the average effect of dust across an extended region.
In this case, the effect of dust is a combination of extinction plus scattering back into the line-of-sight, which implies a strong dependence on the dust geometry, and is referred to as an attenuation curve \citep[see][for a review]{calzetti2001}.
Furthermore, since knowing the dust-free SED shape is crucial for constraining the normalization and wavelength dependence of any extinction or attenuation curve, an additional complication in the case of galaxy attenuation is that, unlike for individual stars, the intrinsic shape of a galaxy SED is less well constrained, being a complex function of a number of unknown parameters (e.g. star-formation history, stellar metallicity, initial mass function etc).

Nevertheless, attenuation curves have been derived for both local starburst galaxies \citep[e.g.][]{calzetti1994,calzetti2000, wild2011,battisti2017} and, more recently, for star-forming galaxies at higher redshifts \citep[e.g.][$z\simeq1.5-3$]{reddy2015}.
These studies have found, for the most part, that the shape of the attenuation curve is similar to the extinction curve, although, due to the different geometry, the increase towards the UV is not as steep, and attenuation curves are often referred to as being greyer than extinction curves in this respect.
A number of studies have attempted to characterize the attenuation curve in terms of physical and structural parameters of galaxies (e.g. inclination, specific star-formation rate), and investigated the presence/absence of the $2175\rm{\AA}$ UV `bump' feature common to many extinction curves, but frequently absent in attenuation curves \citep[e.g.][]{noll2009,conroy2010,buat2011,wild2011,buat2012,kriek2013,salmon2016}.
Nevertheless, by far the most commonly-adopted attenuation law in the galaxy formation research community is the Calzetti starburst law \citep[or Calzetti attenuation curve;][]{calzetti2000}.
For example, our current understanding of the global star-formation rate density and stellar-mass density evolution out to $z\simeq8$ is generally based on the assumption that the Calzetti law is applicable for all star-forming galaxies at all epochs \citep[e.g.][]{madau2014}.

A key question in galaxy evolution is whether, or to what extent, the shape of the galaxy attenuation curve changes, either as a function of galaxy properties (e.g. stellar mass), or as a function of redshift.
Interestingly, a number of recent results in the literature have suggested that the form of the attenuation curve evolves at $z \gtrsim 2$.
For example, several recent sub-mm/mm measurements of the infrared excess (IRX) versus UV spectral slope (i.e. $f_{\lambda} \propto \lambda^{\beta}$) appear to support a scenario in which the attenuation curve becomes much steeper at these redshifts, and more similar in shape to the SMC extinction curve \citep{capak2015,bouwens2016,reddy2017}.
Indeed, some evolution in attenuation properties may be intuitively expected given the growing consensus that star-forming galaxies at $z \gtrsim 2$ are characterised by systematically different physical conditions with respect to their low-redshift counterparts. 
In particular, in combination with an increase of the specific star-formation rates of galaxies with redshift \citep[e.g.][]{esther2016}, an evolution in the physical properties of the interstellar medium (ISM) is also observed, including an increase in ionization parameters and decrease in gas-phase metallicities \citep[e.g][]{cullen2014,kewley2015,strom2017,steidel2016,cullen2016}.
Moreover, at early epochs, possible changes in the relative importance of different dust production channels (e.g. supernova, asymptotic giant branch stars, ISM growth) may become important \citep[e.g.][]{michalowski2010,rowlands2014,watson2015,popping2017}. 
These observations suggest plausible mechanisms (e.g. more extreme radiation fields, different dust production channels) by which the size distribution and composition of dust grains may be different at these epochs.

Nevertheless, there remain a number of uncertainties related to these recent IR measurements which have yet to be fully characterized.
Firstly, IR luminosities carry significant systematic uncertainties related to the unknown dust temperatures in high-redshift galaxies \citep[e.g.][]{schaerer2015,faisst2017}, and biases related to the treatment of $\beta$ may not have been properly accounted for in all cases \citep{mclure2017}.
Secondly, other independent sub-mm/mm observations of the IRX-$\beta$ relation have shown consistency with grey Calzetti-like attenuation curves \citep[e.g.][]{coppin2015,dunlop2017,laporte2017,fudamoto2017,mclure2017}.
To further compound this confusing picture, similarly contradictory results from simulations exist, with some studies supporting a steep UV attenuation curve \citep[e.g.][]{mancini2016} and others finding that Calzetti-like laws are preferable \citep[e.g.][]{cullen2017}
However, since the majority of the inferences regarding the shape of the attenuation curve at $z \gtrsim 3$ remain indirect, one potential avenue of progress is to attempt to directly measure the attenuation curve shape at high redshift.

Predominantly, the method used to derive the attenuation curve is to compare the observed SEDs of a sample of galaxies for which an independent indicator can be used to identify those that are more, or less, affected by dust. 
In practice, the ratio of hydrogen emission lines (most commonly \halpha/\hbeta) is used to rank galaxies in increasing order of their Balmer decrement \citep{calzetti1994, reddy2015}\footnote{Since this involves the ratio of nebular emission lines, a separate connection must be made to prove that the increasing attenuation of the ionized gas corresponds to increasing attenuation of the stellar continuum. 
Commonly, to provide this connection, the Balmer optical depth is shown to correlate with the observed slope of the UV continuum \citep[e.g][]{calzetti1994}}.
The primary advantage of this approach is that, in not having to assume the shape of a dust-free SED, potentially large systematic errors are avoided.
However, one disadvantage is that since the dust-free intrinsic SED shape is not known, this method is sensitive to the total-to-selective attenuation law, rather than the attenuation law directly.
An alternative approach is to explicitly assume the underlying intrinsic SED of all sources.
In this case it is possible to directly probe the wavelength dependence of the attenuation curve by taking the ratio of intrinsic to observed spectra \citep[e.g.][]{scoville2015}.
This method suffers from an increase in systematic errors related to the choice of the intrinsic SED shape. However, these problems will be mitigated somewhat at high redshifts where star-formation histories are predicted to become self similar across the galaxy population \citep[e.g.][]{finlator2011}.

    \begin{figure}
        \centerline{\includegraphics[width=\columnwidth]{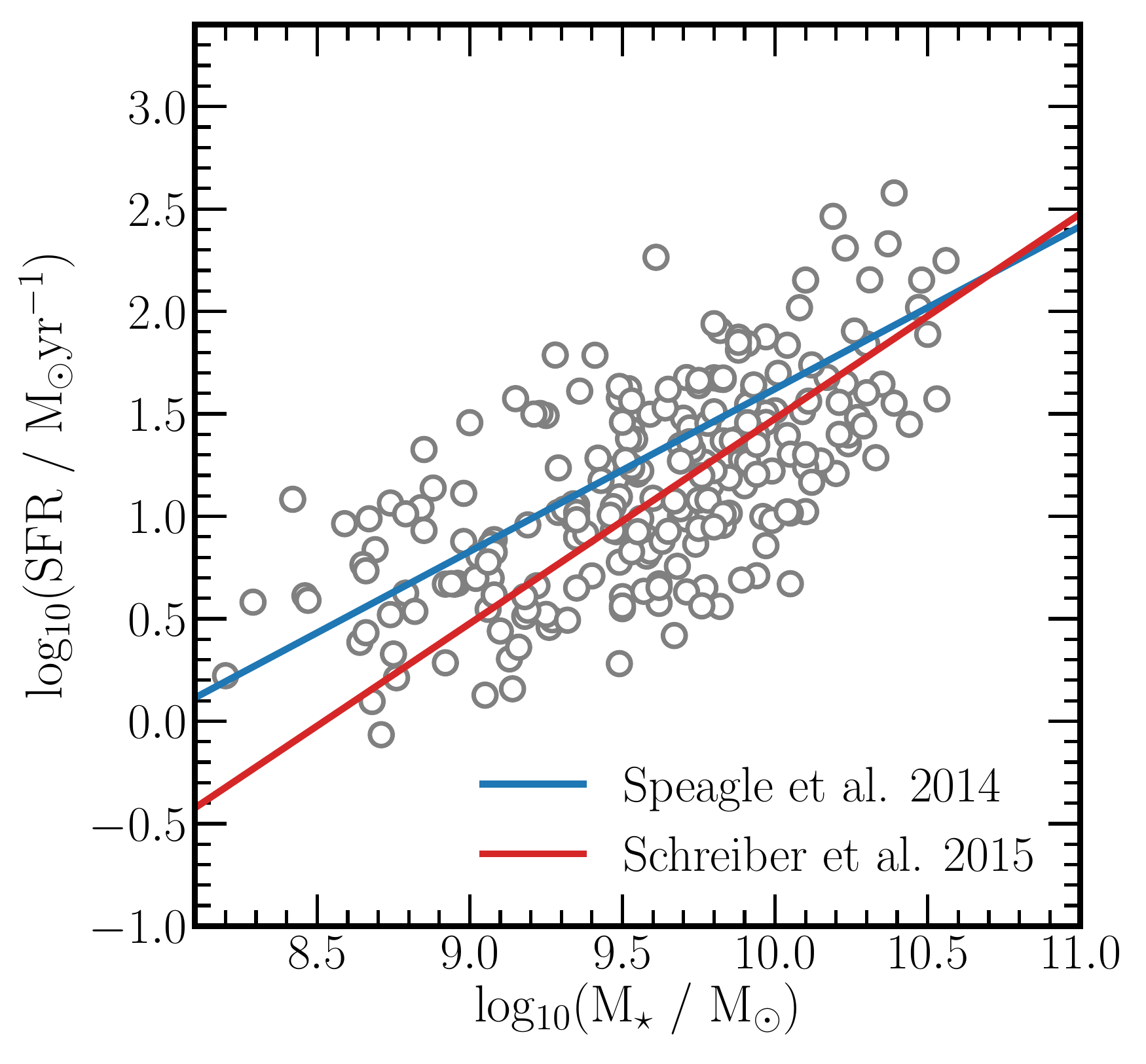}}
        \caption{The position of our VANDELS sample in the M$_{\star}$-SFR plane.
        Masses and star-formation rates were derived using the SED fitting code \textsc{LePahre} (see text for details).
        For comparison, the blue and red lines show two independent determinations of the main sequence of star formation at $z=3.5$ from \citet{speagle2014} and \citet{schreiber2015} respectively.
        It can be seen that the galaxies in our sample are consistent with being typical star-forming galaxies at $z=3.5$.}
        \label{fig_main_sequence}
    \end{figure}

In this paper we implement this type of direct analysis to derive the shape of the attenuation curve for a preliminary sample of $236$ galaxies at $3 < z < 4$, selected from the VANDELS survey \citep{vandels_messenger}.
Crucially, building on the study of \citet{scoville2015}, we construct a set of intrinsic, dust-free, SED templates that contains both simple constant star-formation rate models as well as physically-motivated SEDs derived from the First Billion Years (FiBY) simulation \citep{paardekooper2015}. 
Furthermore, we utilise a comparable sample size over a much narrower redshift range, mitigating any redshift-dependent systematics.
Our aim is to investigate whether we find any evidence to suggest that the average shape of the attenuation curve is evolving with redshift, or whether the commonly-adopted attenuation curve of \citet{calzetti2000} is still, on average, applicable for star-forming galaxies at $z\simeq3.5$.

The structure of the paper is as follows.
In Section \ref{sec_data} we describe the VANDELS data and the ancillary photometry used in our analysis, and detail the assumptions and methods used to generate the intrinsic galaxy SEDs.
In Section \ref{sec_attn_curve_derivation} we outline our method for deriving the attenuation curve before presenting our results in Section \ref{sec_results}, including a investigation of the mass-dependence of both the total attenuation and the shape of the attenuation curve.
Finally, we summarize our conclusions in Section \ref{sec_conclusions}.
We adopt the following cosmological parameters throughout this paper: $\Omega_m =0.3$, $\Omega_\Lambda =0.7$, $H_0 =70$ km s$^{-1}$ Mpc$^{-1}$.
All stellar masses are calculated using a \citet{chabrier2003} initial mass function (IMF).

\section{Data and Models}\label{sec_data}

In this section we provide a brief overview of the spectroscopic and imaging datasets used in the current study and define the final sample of star-forming galaxies.  
We also provide a full description of how the observed SEDs were derived from the rest-frame optical photometry, and outline how the intrinsic template SEDs were generated.

\subsection{VANDELS Survey}

VANDELS is a public ESO spectroscopic survey targeting $\rm{N} \approx 2000$ galaxies at $1.0 < z < 7.0$ with the VIMOS spectrograph on the VLT in the CDFS and UDS extragalactic survey fields.
Observations have been performed using the medium resolution grism, which provides a spectral resolution of $R \approx 600$ over the wavelength range $4800-10000 \rm{\AA}$ at a dispersion of $2.5 \rm{\AA}$ per pixel.
Details of the survey and data reduction will be provided in two upcoming papers (McLure et al. in prep; Pentericci et al. in prep).
The survey is due to be completed in January 2018, therefore in this work we use the current sample of $1502$ fully or partially observed spectra, from which we select $242$ galaxies with secure redshifts in the range $3.0\leq z \leq 4.0$ (see Section \ref{sec_final_vsamp} for more details).
Galaxies at the redshift of our sample are selected to have $I_{AB} \leq 27.5$ and $H_{AB} \leq 27$, with integration times optimized to provide 1D spectra with a signal-to-noise-ratio (SNR) of 15 - 20 per resolution element.
Since not all the spectra used this paper have been fully observed, this SNR is not achieved for all galaxies in our sample.

\subsection{Photometry}

    \begin{figure}
        \centerline{\includegraphics[width=\columnwidth]{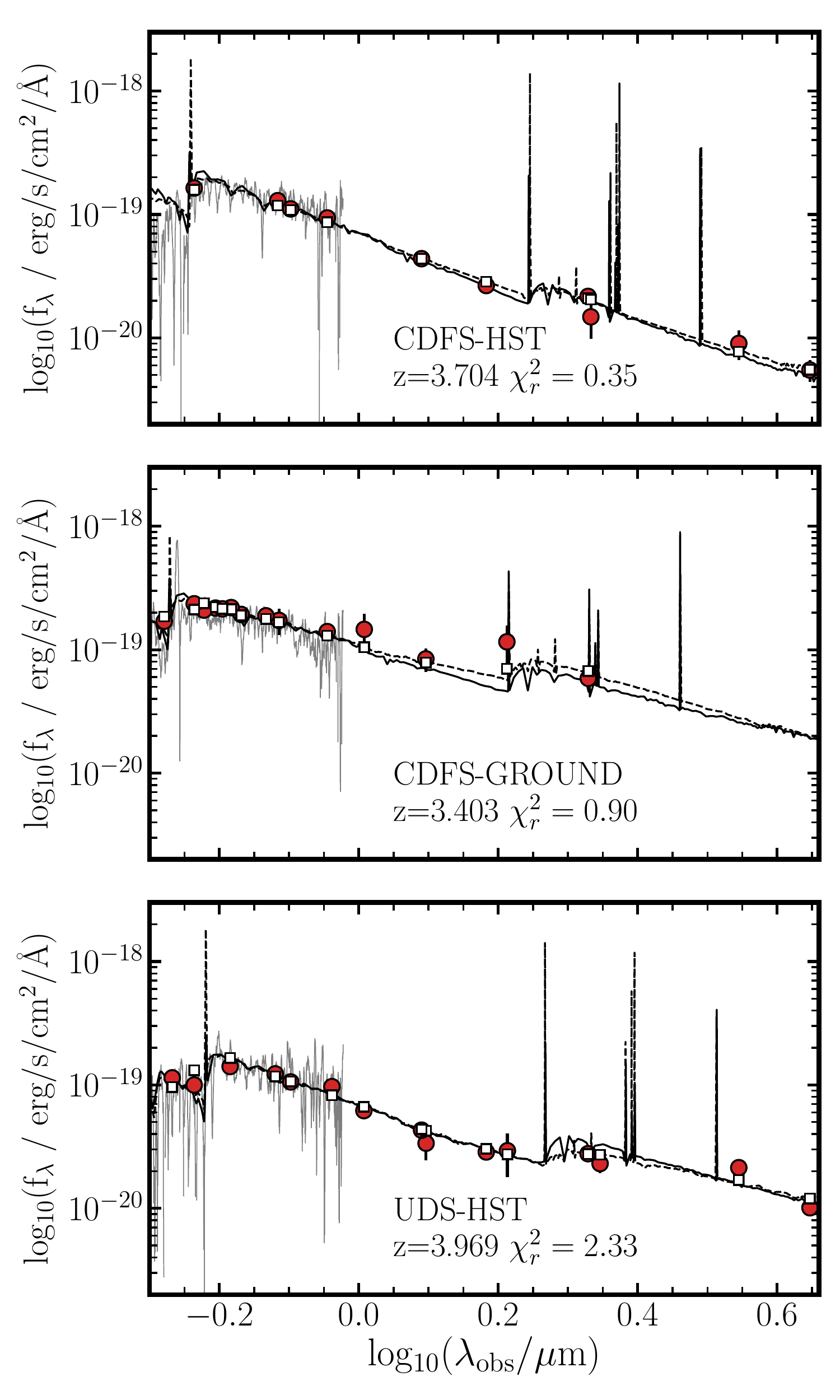}}
        \caption{The UV-to-optical spectral energy distributions for three galaxies in our sample.
        In each panel, the red circular data points show the observed photometry while black solid/dashed spectra show the best-fitting LePhare/EAZY templates respectively.
        The open squares show the synthetic photometry of the best-fitting EAZY template through the observed filters.
        The grey spectra at $\lambda \approx 0.5 - 1.0$ $\mu \rm{m}$ ($\mathrm{log}(\lambda / \mu\mathrm{m}) \approx -0.3 - 0.0$) are overlays of the observed VANDELS spectra, which we used to accurately determine the redshift of the galaxies.}
        \label{fig_sed_fitting}
    \end{figure}

All VANDELS targets have been selected from the CDFS and UDS survey fields.
Within each field, $\simeq 50 \%$ of galaxies fall within the CANDELS region \citep{grogin2011,koekemoer2011} and therefore benefit from deep optical to near-infrared \em Hubble Space Telescope \em (HST) imaging, while the other $50 \%$ fall within the wider survey regions and relies on predominantly ground-based imaging.
For the remainder of this paper we reference these four separate subsets by their field (CDFS or UDS) and whether they are covered by primarily space-based (HST) or ground-based (GROUND) imaging (i.e. CDFS-HST, CDFS-GROUND, UDS-HST and UDS-GROUND).

The CDFS-HST photometry utilizes most of the same filters as the publicly-available CANDELS catalog described in \citet{guo2013}, with the difference that the $Ks$ HAWKI photometry has been updated to include the final HUGS survey data \citep{fontana2014} and the CTIO U-band filter has been removed.
The final catalog contains photometry across the wavelength range $0.3-8.0 \mu \rm{m}$ (i.e. covers the rest-frame UV to near-infrared SEDs of our sample out to $\lambda_{rest} \sim 2.0 \mu \rm{m}$).
Similarly, the UDS-HST photometry is essentially identical to the public CANDELS catalog published by \citet{galametz2013}, and covers the same wavelength range.
The CDFS-GROUND and UDS-GROUND photometry, whilst including some HST filters, are both composed primarily of ground-based observations.
Crucially, these regions are not covered by deep Spitzer/IRAC observations, so the observed photometry only extends out the K-band ($\sim 2.2 \mu \rm{m}$) and the rest-frame photometry therefore only covers the UV to optical SED of our sample out to $\lambda_{rest} \sim 0.6 \mu \rm{m}$.
In the attenuation curve analysis presented in this paper, we confine our analysis to the $0.1 \lesssim \lambda \lesssim 0.6 \mu \rm{m}$ region.
A detailed description of the various photometric catalogues will be given in the VANDELS survey definition paper (McLure et al. in prep).

\subsection{VANDELS Sample}\label{sec_final_vsamp}

The sample analysed in this paper is drawn from the current batch of $1502$ VANDELS spectra.
Of these, $480$ have formal VANDELS redshifts in the range $3.0 \leq z \leq 4.0$, however for this initial study we restrict this sample to only those with the most secure redshift flags (FLAG = 3 or 4, i.e. a $> 95 \%$ probability of being correct) to avoid any potential redshift contamination issues (see McLure et al. in prep).
This restricted sample contains $242$ galaxies.
We then applied a mass cut of  log$(M_{\star}/M_{\odot}) \leq 10.6$, leaving a total of $236$ galaxies.
As described in Section \ref{sec_intrinsic_seds}, throughout this paper we compare our observed sample to a sample of simulated galaxies from the FiBY simulation, which only contains galaxies with stellar masses up to log$(M_{\star}/M_{\odot})=10.6$, so we have applied the same upper mass limit to the observed sample.
A description of how stellar masses were derived from the observations is given below.
Therefore, the final selection criteria for the sample used in this paper is simply the original VANDELS selection criteria ($I_{AB} \leq 27.5$ and $H_{AB} \leq 27$ with a photometric redshift solution in the correct redshift range) plus the requirement of a secure spectroscopic redshift from current observations and an upper mass limit of log$(M_{\star}/M_{\odot})=10.6$.
Fig. \ref{fig_main_sequence} illustrates how the effect of these criteria is to select galaxies consistent with the star-forming main sequence at $3.0 < z < 4.0$ (see also McLure et al. in prep).
The galaxies in our final sample span stellar masses in the range $1.6 \times 10^8 \leq M_{\star}/M_{\odot} < 4.0 \times 10^{10}$ with a median value of $4.5 \times 10^9$ $\rm{M}_{\odot}$, at a mean redshift of $\langle z \rangle = 3.49$.

\subsection{Observed SEDs}

    \begin{figure}
        \centerline{\includegraphics[width=\columnwidth]{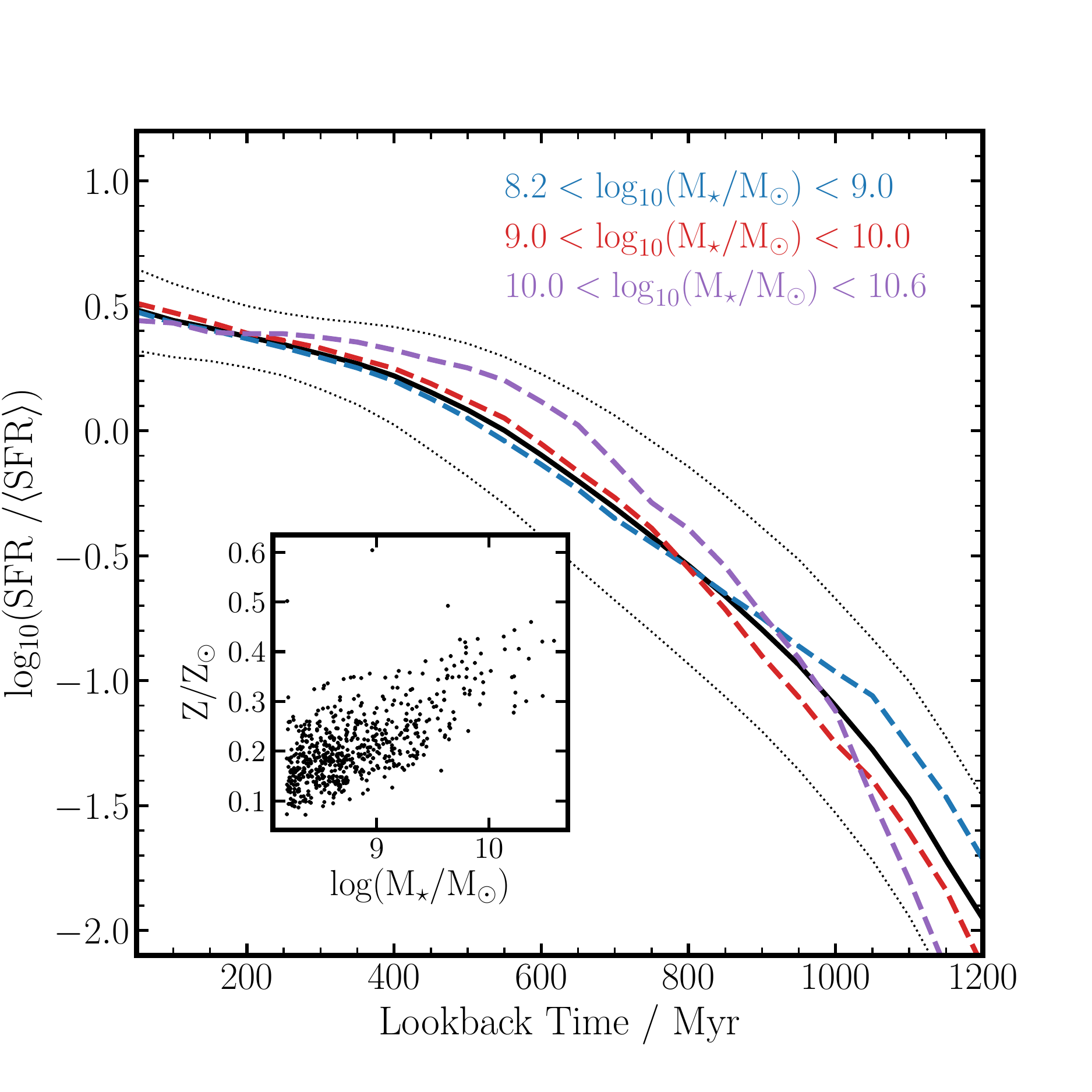}}
        \caption{The average normalised star-formation histories and metallicities of $628$ simulated galaxies extracted from the FiBY simulation at $z=4$, covering the stellar mass range $8.2 \leq $ log $(M_{\star}/M_{\odot}) \leq 10.6$.
        The black solid curve shows the mean normalised star-formation history with the black dotted curve representing the $\pm 1 \sigma$ scatter.
        The vast majority of the galaxies have self-similar, rising, star-formation histories with an average UV-weighted metallicity of $\langle Z/Z_{\odot} \rangle = 0.21 \pm 0.07$ (inset panel).
        The consistency of star-formation histories is further illustrated by the three colored dashed curves, which show the mean star-formation history in three stellar mass bins, with the corresponding stellar mass ranges given in the upper right-hand corner.
        As discussed in the text, this conformity of star-formation history results in the intrinsic UV to optical SED shape of all galaxies being consistent to within $\approx 10\%$ (averaged across all wavelengths).
        This is the basis of our assumption that the observed SED shapes can be compared to a single common intrinsic SED shape, allowing us to directly probe the attenuation curve.
        }
        \label{fig_sfhs}
    \end{figure}

To determine the observed SED shape of each galaxy we used two publicly-available SED-fitting codes: LePhare \citep[e.g][]{ilbert2008} and EAZY \citep{brammer2008}.
EAZY is a redshift fitting code which uses linear combinations of a limited standard template set to best match the observed photometry. 
The template set is derived from fitting synthetic photometry of a semianalytic model (SAM) with the PEGASE stellar population synthesis library \citep{fioc1997}.
These standard templates can be thought of as the `principal component' templates of all SED shapes in the SAM over the redshift range $0 < z \lesssim 4$ \citep{brammer2008}.
We ran EAZY on all galaxies in our sample using the default parameter set, including emission lines, fixing the redshift to the spectroscopic redshift measured from the VANDELS spectra.
A major benefit of EAZY is the relatively small number of \emph{a-priori} assumptions required; for example is it not necessary to assume a dust attenuation curve or star-formation history.
Examples of fits to three galaxies are shown in Fig. \ref{fig_sed_fitting}.

On the other hand, LePhare allows one to fit the observed photometry using a range of SEDs built using a chosen stellar-population synthesis model library along with a parameterized star-formation history, dust attenuation law, colour excess and metallicity.
The main advantage of this approach is that the stellar mass of the galaxy can be reliably estimated, which we used in building the intrinsic template set, investigating the attenuation versus stellar-mass relation, and in splitting the sample by mass.
To run LePahre we built the SED library using constant star formation BC03 stellar population synthesis models \citep{bruzual2003} at three metallicities (0.28, 0.56 and 1.4 Z$_{\odot}$)\footnote{Through this paper we assume $Z_{\odot}$=0.0142 \citep{asplund2009}. These metallicity values approximately cover the range of metallicities recovered from the FiBY simulation at $z=4$ (see Section \ref{sec_intrinsic_seds}).} with 0.0 $\leq$ E(B$-$V) $\leq$ 0.4 assuming the \citet{calzetti2000} attenuation curve.
Again, examples of the resulting fits are shown in Fig. \ref{fig_sed_fitting}.
To check that our assumption of a constant star-formation history was not biasing the resulting stellar masses, we also ran a set of exponentially rising star-formation history models with $e-$folding times $\tau = 50-10000$ Myr using the Bayesian SED fitting code \textsc{bagpipes} \citep{carnall2017}, adopting the same set of SPS models, metallicities and color excesses.
The masses and star-formation rates returned from this test are fully consistent with the data in Fig. \ref{fig_main_sequence}.
We note that, given that the shapes of the SEDs are well constrained by the available photometry, the actual differences in the shapes of the resulting best-fitting templates are small and insensitive to model assumptions.
Indeed, from  Fig. \ref{fig_sed_fitting} it can be seen that similar SED shapes would be obtained by a simple linear interpolation between the SED points.
The median reduced $\chi^2$ values for the LePhare and EAZY fits are 1.48 and 1.50 respectively; for the final observed SED of each galaxy, we averaged the best-fitting LePhare and EAZY SEDs.

\subsection{Intrinsic template SED}\label{sec_intrinsic_seds}

As we describe in detail in Section \ref{sec_attn_curve_derivation}, to derive the attenuation curve, the observed SED shape needs to be compared to an estimate of the intrinsic SED shape.
In this paper, we make the assumption that all galaxies in the sample have the same intrinsic UV to optical SED shape.

\subsubsection{Evidence for a homogeneous SED shape}

    \begin{figure}
        \centerline{\includegraphics[width=\columnwidth]{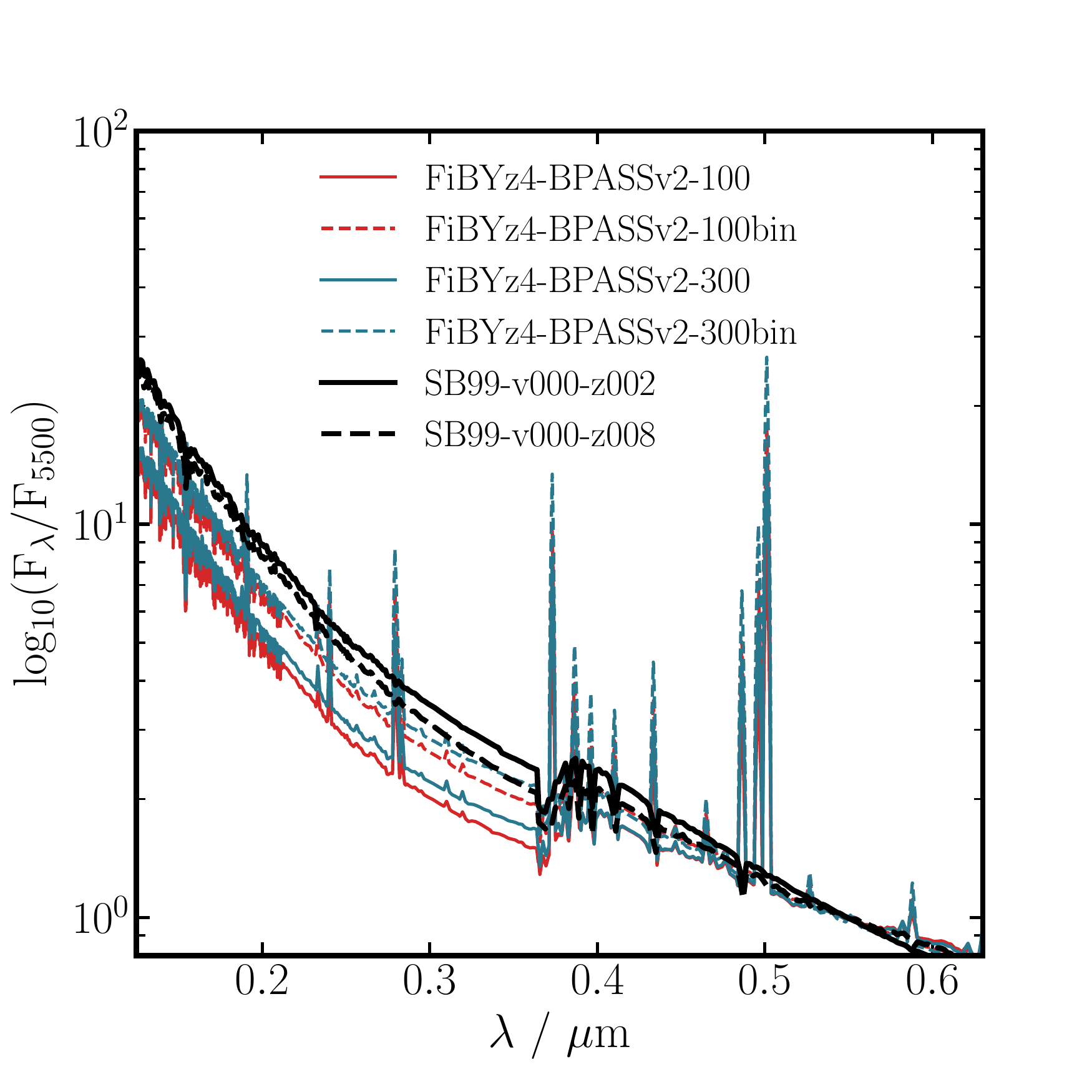}}
        \caption{The six template spectra used as the intrinsic (dust-free) SED shapes when deriving the dust attenuation curve.
        The red lines show stacked synthetic spectra of FiBY simulated galaxies at $z=4$ (see text for details) for the FiBY-BPASSv2-100 model (solid) and FiBY-BPASSv2-100bin model (dashed) respectively.
        The blue lines show the same for the BPASSv2 models with an upper IMF cutoff of $300 M_{\odot}$ (FiBY-BPASSv2-300, FiBY-BPASSv2-300bin).
        Finally, the solid and dashed black lines show the S99-v00-z002 and S99-v00-z008 constant star-formation rate Starburst99 models, both with age 100 Myr, respectively.
        All spectra are normalised at $0.55 \mu\rm{m}$.}
        \label{fig_intrinsic_templates}
    \end{figure}

We were able to assess the validity of this assumption using data from the FiBY cosmological hydrodynamical simulation. 
A sample of $628$ galaxies, with $8.2 \leq \rm{log(M/M_{\odot})} \leq 10.6$ (matching the mass range of the observed sample) at $z=4$ (the minimum redshift of the simulation), was extracted from two simulation boxes with a combined comoving volume of $\approx 3.7 \times 10^4 \rm{Mpc}^3$ \citep[see][for simulation details]{johnson2013,cullen2017}.
Fig. \ref{fig_sfhs} shows the normalised mean star-formation history and $\pm 1 \sigma$ scatter for the full sample of simulated galaxies, as well as the mean star-formation histories split into three $M_{\star}$ bins.
To construct Fig. \ref{fig_sfhs} we first normalised the star-formation history of each galaxy using its mean star-formation rate, which is the total stellar mass formed at $z=4$ divided by the age of the oldest star particle in the galaxy (i.e. an approximation of the formation age of the galaxy).
Fig. \ref{fig_sfhs} illustrates how the typical star-formation histories are self-similar across the whole population with a relatively small scatter.
For example, over a lookback time of $\sim 300$ Myr (from $z=4$), where star-formation rates are typically $3 \times$ the mean, the scatter is $\sim 0.15 $ dex (a factor $\sim 1.4$).
Furthermore, in the inset panel of Fig. \ref{fig_sfhs} we show the UV-weighted metallicity of the galaxies (an approximation of the mean metallicity of the galaxies formed over the last 100 Myr) as a function of stellar mass.
Again, the distribution is relatively tight, with a mean UV-weighted metallicity of $\langle Z/Z_{\odot} \rangle_{\rm{Fe}} = 0.21 \pm 0.07$.
Based on the similarity of the star-formation histories, and the narrow range in metallicity, it is reasonable to assume that the underlying SED shape will be similar across all masses.
In Section \ref{sec_av_mass} we provide further empirical evidence for the validity of this assumption.

    \begin{figure*}
        \centerline{\includegraphics[width=7in]{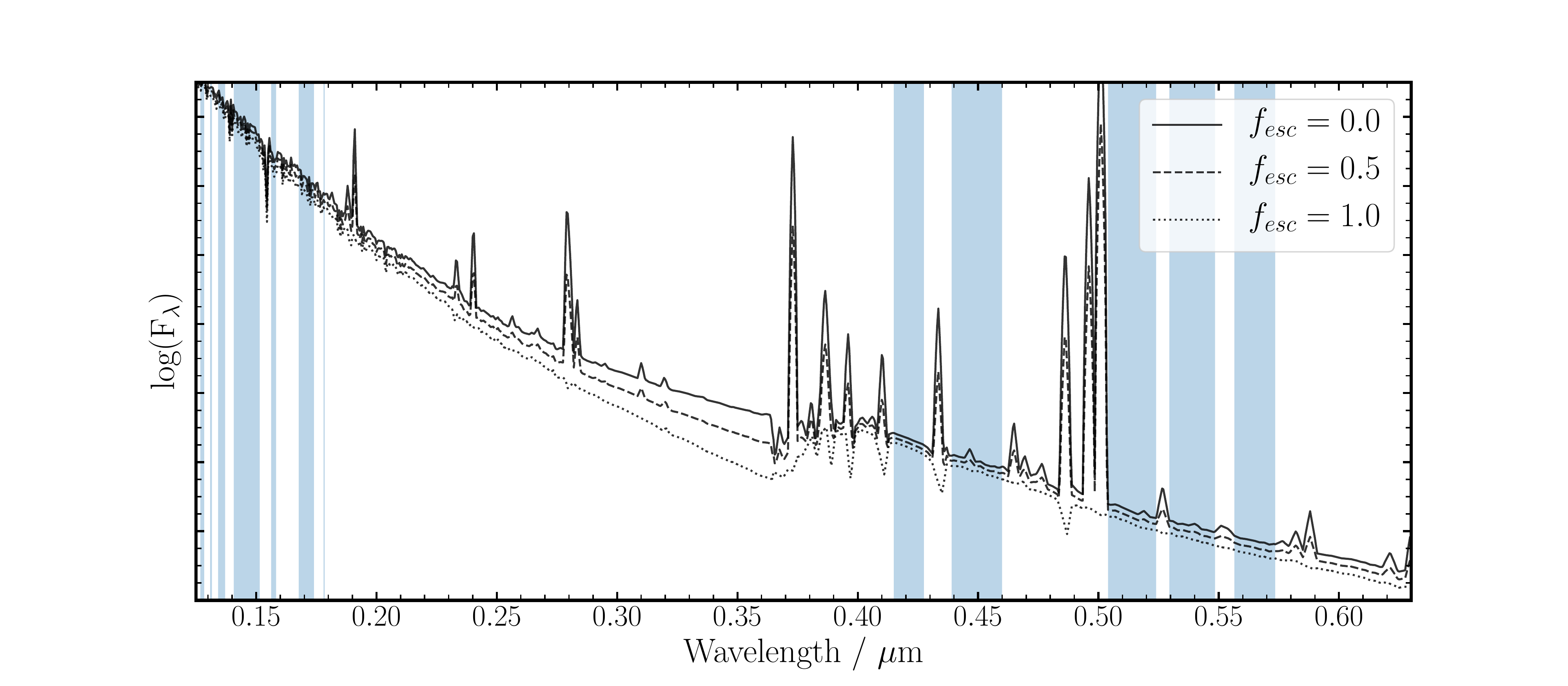}}
        \caption{The intrinsic FiBY-BPASSv2-300bin SED across the wavelength region $0.125 \mu \rm{m} < \lambda < 0.63 \mu \rm{m}$ assuming different values for the escape fraction of ionizing photons ($f_{esc}$).
        It can be seen that some wavelength regions are extremely sensitive to the nebular spectrum and therefore unsuitable to use when fitting the attenuation curve (see text for details).
        The blue shaded areas show the wavelength regions we defined as suitable to use. 
        These wavelength pixels satisfied the following criteria: (i) a maximum nebular contribution of $10\%$; (ii) free from any strong stellar absorption/emission features.}
        \label{fig_fitting_regions}
    \end{figure*}

It is important to note here that we are defining metallicity as the iron abundance relative to solar ($[\rm{Fe}/\rm{H}]$) rather than the total abundance of all metals.
This is motivated by the fact that stellar opacity in the UV, and hence the UV spectral shape, is dominated by Fe and is relatively insensitive to $[\rm{O}/\rm{H}]$ \citep[e.g.][]{rix2004}. 
This is important because, for galaxies with rising star-formation histories, $[\rm{Fe}/\rm{H}] \neq [\rm{O}/\rm{H}]$ \citep[e.g.][]{steidel2016, cullen2017}.
This can be seen in the FiBY sample, where the mean UV-weighted total metallicity of all 628 simulated galaxies is $\langle Z/Z_{\odot} \rangle_{\rm{Tot}} = 0.76 \pm 0.30$ (i.e. a factor $\approx 3.5$ larger than the Fe metallicity).
As a consequence, since most standard SPS models assume solar abundance ratios, in practice one must choose which of the two abundance definitions to use as the `true' metallicity when constructing SEDs from SPS models.
For this paper we adopt the UV-weighted Fe abundances for constructing the SEDs.

   \begin{table}
        \centering
        \caption{A list of all the templates used for fitting the observed SED shapes, and generating the intrinsic SED shapes.}\label{tab_templates}
        \begin{tabular}{llc}
            \hline
            Type & Templates & $\beta^a$ \\
            \hline
            \hline
            Observed & EAZY (Default template set, PEGASE) & -- \\ 
                     & LePhare (BC03 CSF) & -- \\
            \hline
            \hline
            Intrinsic & FiBY-BPASSv2-100bin$^b$ & -2.35 \\
                      & FiBY-BPASSv2-300bin$^b$ & -2.31 \\
                      & FiBY-BPASSv2-100$^c$ & -2.34 \\
                      & FiBY-BPASSv2-300$^c$ & -2.31 \\
                      & S99-v00-z002$^d$ & -2.41\\
                      & S99-v00-z008$^d$ & -2.41\\
            \hline
            \multicolumn{2}{l}{$^a$ UV continuum slopes of the intrinsic templates}\\
            \multicolumn{2}{l}{$^b$ FiBY binary star models}\\
            \multicolumn{2}{l}{$^c$ FiBY single star models}\\
            \multicolumn{2}{l}{$^d$ Starburst99 models}\\
        \end{tabular}
    \end{table}

\subsubsection{Building the intrinsic SEDs}

The simulated galaxy SEDs from FiBY were built as described in \citet{cullen2017}.
Briefly, each star particle associated with a given galaxy was assigned an instantaneous starburst stellar population synthesis model which best matched its age and metallicity, and the final galaxy SED was then constructed by summing up the SEDs of all individual star particles.
For the stellar population synthesis models we used BPASSv2 \citep[e.g.][]{eldridge2016,stanway2016}, and considered their four fiducial models, which we refer to according to the upper mass cutoff of the IMF and whether or not binary evolution is included.
The BPASSv2-100bin models include binary star evolution with an IMF cutoff of 100 M$_{\odot}$, and BPASSv2-100 are the equivalent single-star evolution models; similarly, BPASSv2-300bin models include binary evolution with an IMF cutoff of 300 M$_{\odot}$ and BPASSv2-300 are the equivalent single-star evolution models.
All models have an IMF index of $-1.3$ between $0.1 - 0.5$ M$_{\odot}$ and $-2.35$ above 0.5 M$_{\odot}$.
Finally, the nebular continuum contribution is included using \textsc{cloudy} \citep{ferland2017} as described in \citet{cullen2017}, assuming maximal nebular contribution (i.e. assuming escape fraction $f_{esc}=0\%$).

To generate the set of intrinsic template SEDs used in our analysis we constructed, for each of the four BPASS models, a stack of FiBY SEDs over a similar mass range to the observed sample.
To generate each stack, we paired each VANDELS spectrum with the FiBY galaxy which matched it closest in mass, avoiding duplications.
Thus, the mass distribution of the stacked FiBY SED was the closest possible match to the mass distribution of the VANDELS sample.
Fig. \ref{fig_intrinsic_templates} shows the stacked SEDs for all FiBY models. 
For a given stack (e.g. FiBY-BPASSv2-100bin) the standard deviation in each pixel, across the wavelengths of interest ($0.125 \mu \rm{m} \leq \lambda \leq 0.63 \mu \rm{m})$, ranges from $\approx 1 - 20 \%$ with an average of $9\%$. 
We note that, given the similarity in star-formation history and metallicity across all masses, this method of mass-matched stacking yielded very similar results to simply stacking all 628 FiBY SEDs.

We also considered two constant star-formation rate models from the latest version of Starburst99 \citep{leitherer2014} assuming the weaker-wind Geneva tracks without stellar rotation, a \citet{kroupa2001} IMF and an age of 100 Myr.
The two models had metallicities $Z_*=0.002$ and $Z_*=0.008$ respectively.
These models assume single-star evolution and an upper-mass IMF limit of $100\rm{M}_{\odot}$, include nebular-continuum emission, and are referred to as S99-v00-z002 and S99-v00-z008 (see Table \ref{tab_templates}).
The $Z_*=0.002$ ($\rm{Z/Z_{\odot}} = 0.14$) model was chosen as the closet match to the mean UV-weighted \rm{Fe}/\rm{H} based metallicity of the simulated sample ($\langle \rm{[Fe/H]} \rangle = 0.20$) while the $Z_*=0.008$ ($\rm{Z/Z_{\odot}} = 0.56$) was chosen as the closest match to the mean UV-weighted total metallicity ($\langle \rm{Z/Z_{\odot}} \rangle = 0.70$).
Comparing the two Starburst99 models allowed us to test how sensitive our results are to the assumed metallicity.
Finally, an age of 100 Myr was chosen to match the typical UV and optical-weighted ages of the simulated galaxies.
The mean ages weighted by $0.15$ $\mu\rm{m}$ and $0.55$ $\mu\rm{m}$ luminosity are $44\pm5$ Myr and $150\pm6$ Myr for the FiBY sample (averaged across the four BPASSv2 models).
An age of 100 Myr is therefore a reasonable compromise between the estimated age of the stellar populations dominating the UV and the optical luminosity at $z\simeq3.5$.

Fig. \ref{fig_intrinsic_templates} shows the six intrinsic SED templates described above normalised at $0.55 \mu \rm{m}$.
A summary of all of the observed and intrinsic template sets is given in Table \ref{tab_templates}.

\section{Deriving the attenuation law}\label{sec_attn_curve_derivation}

In this section we describe the method used to derive the shape and normalization of the UV $-$ optical attenuation curve and assess the validity of the method using a simple simulation

\subsection{Fitting method}

   \begin{table}
        \centering
        \caption{The 12 rest-frame wavelength regions used for fitting the attenuation curve (shown in Fig. \ref{fig_fitting_regions}).}\label{tab_fit_windows}
        \begin{tabular}{cc}
            \hline
            \hline
            Window Number & Wavelength range ($\mu \rm{m}$) \\
            \hline
            1 & $0.1268 - 0.1284$ \\
            2 & $0.1309 - 0.1316$ \\
            3 & $0.1342 - 0.1371$ \\
            4 & $0.1407 - 0.1510$ \\
            5 & $0.1562 - 0.1583$ \\
            6 & $0.1677 - 0.1740$ \\
            7 & $0.1780 - 0.1785$ \\
            8 & $0.4150 - 0.4275$ \\
            9 & $0.4390 - 0.4600$ \\
            10 & $0.5040 - 0.5240$ \\
            11 & $0.5295 - 0.5485$ \\
            12 & $0.5565 - 0.5735$ \\
            \hline
        \end{tabular}
    \end{table}

    \begin{figure*}
        \centerline{\includegraphics[width=7in]{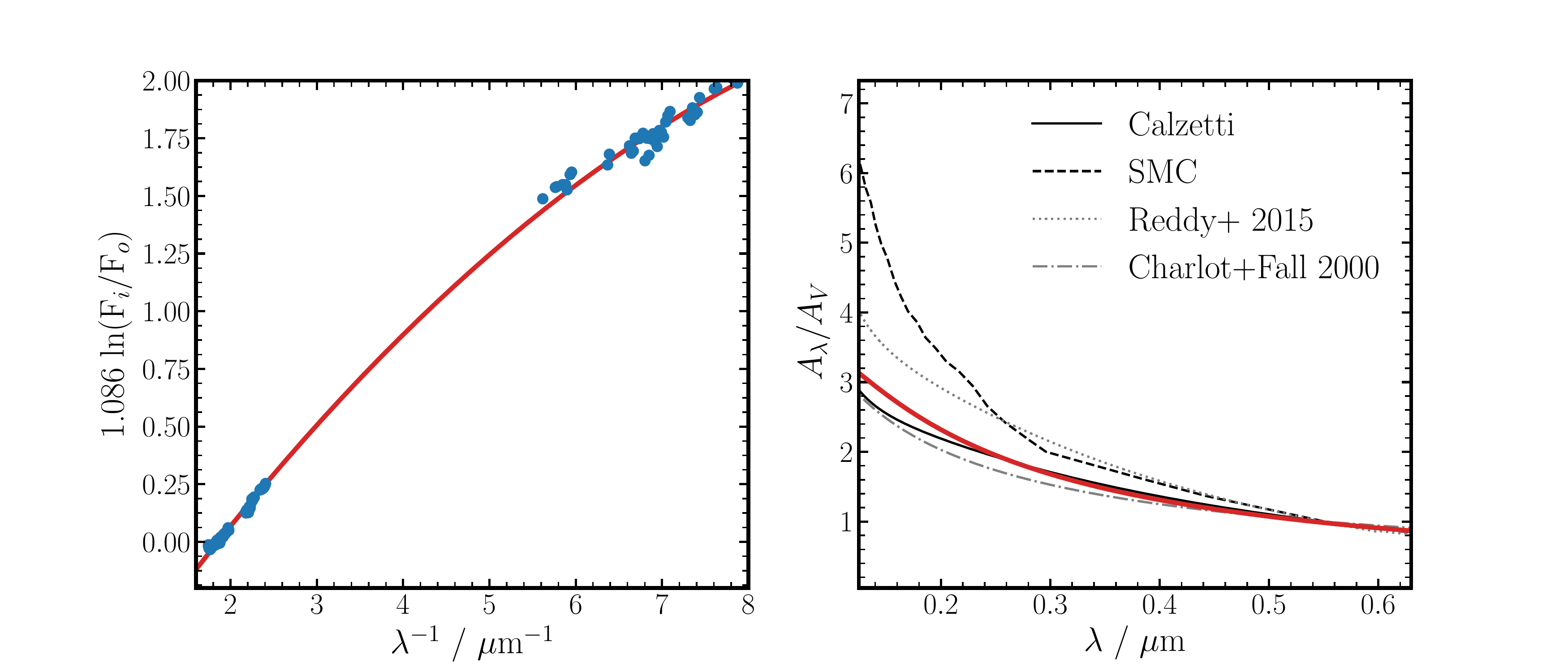}}
        \caption{
        The left-hand panel shows an attenuation curve fit for one galaxy in our sample as described in Section \ref{sec_attn_curve_derivation}, with the wavelength pixels used for the fitting shown as the blue dots (see text for discussion).
        The final best-fitting solution is shown by the solid red curve.
        The resulting fit, translated into $A_{\lambda}/A_V$ versus $\lambda$ is shown in the right-hand panel and is compared to the corresponding curves for the \citet{calzetti2000}, \citet{reddy2015} and \citet{charlot2000} attenuation curves, and the SMC extinction curve of \citet{gordon2003}.
        For this particular galaxy the solution suggests a grey, Calzetti-like, shape.}
        \label{fig_example_fitting}
    \end{figure*}

If the observed and intrinsic SED shapes are known, then the equation relating them is
\begin{equation}\label{eq_dust_optical_depth}
F_{\lambda,o} = \phi F_{\lambda,i} e^{-\tau_{\lambda}},
\end{equation}
where $\phi$ is the unknown normalization of the intrinsic SED, and $\tau_{\lambda}$ is the wavelength-dependent optical depth.
Evaluating at an arbitrary wavelength, which we take as $0.55 \mu \rm{m}$ (and refer to using the subscript $V$) we can solve for $\phi$, and, after converting to $\rm{A}_{\lambda}$ using $A_{\lambda} = 1.086 \tau_{\lambda}$, Equation \ref{eq_dust_optical_depth} can be rearranged to give
\begin{equation}\label{eq_dust_final}
 1.086 \times \mathrm{ln}(F_{\lambda,i}/F_{\lambda,o}) =  A_{\lambda} - A_{V}.
\end{equation}
where $F_{\lambda,i}$ and $F_{\lambda,o}$ are both normalised at $\lambda=0.55 \mu \rm{m}$.
For $A_{\lambda}$ we adopt a second-order polynomial as a function of $1/\lambda$, and set the additive constant term to zero such that $A_{\lambda} \to 0$ as $\lambda \to \infty$:
\begin{equation}\label{eq_neta}
A_{\lambda} = a_{0}x + a_{1}x^2,
\end{equation}
where $x=1/\lambda$ $\mu\rm{m}^{-1}$.
We discuss this choice of parameterization further in Section \ref{sec_attn_curve_sims}.
Therefore, by fitting
\begin{equation}\label{eq_dust_final}
 1.086 \times \mathrm{ln}(F_{\lambda,i}/F_{\lambda,o}) =  a_{0}x + a_{1}x^2 - A_{V},
\end{equation} 
one can derive the shape and normalization of the attenuation curve of a given galaxy.

Finally, the more commonly adopted parameterizations of the attenuation curve are readily obtained via
\begin{equation}\label{eq_rv}
R_{V} = \frac{A_{V}}{(A_{B} - A_{V})},
\end{equation}
where $R_V$ is referred to as the total-to-selective attenuation ratio and $A_{B}$ is the attenuation at $\lambda=0.44 \mu \rm{m}$, and
\begin{equation}\label{eq_klam}
k_{\lambda} = \frac{A_\lambda R_V}{A_V},
\end{equation}
where $k_{\lambda}$ is the total-to-selective attenuation curve.

The first step was to re-sample the observed and intrinsic SED templates onto a common rest-frame wavelength grid.
We adopted a wavelength grid spanning $0.12 - 0.63 \mu\rm{m}$ with steps $d\lambda=5\times10^{-4} \mu \rm{m}$ ($5\rm{\AA}$).
The upper and lower limits were chosen to mimic the wavelength window used in the derivation of \citet{calzetti2000} attenuation curve at UV to optical wavelengths. 
Moreover, $0.63 \mu \rm{m}$ is a reasonable upper limit since it is close to the maximum rest-frame wavelength at which all galaxies in our sample have photometric coverage.

The second step was to decide which wavelength pixels to use when performing the fit.
Not all pixels are suitable since some will be biased by strong stellar and nebular features.
We adopted the following criteria to select good pixels: (i) those with an estimated nebular contribution (line or continuum) of $< 10 \%$; (ii) those free from stellar absorption/emission features.
To implement (ii) we used the \citet{calzetti1994} windows in the UV, and a set of custom windows at optical wavelengths masking the strong hydrogen absorption features.
The resulting rest-wavelength regions that we used for fitting the attenuation curve are listed in Table \ref{tab_fit_windows} and shown in Fig. \ref{fig_fitting_regions}. 
An example fit to one galaxy in our sample is shown in Fig. \ref{fig_example_fitting}.

\subsubsection{The nebular continuum and emission spectrum}

We decided to exclude regions of the spectrum with a significant nebular contribution due to the large uncertainties inherent in modeling the nebular emission.
In particular, the nebular spectrum is sensitive to the assumed escape fraction ($f_{esc}$) of ionizing photons, which depends primarily on the covering fraction of neutral hydrogen and the dust content of a galaxy \citep[e.g.][]{hayes2011}.

Our nebular modeling is based on the latest determinations of physical parameters within \hii \ regions at $z\simeq 2.5$ \citep[e.g.][]{steidel2016, strom2017} and is described in detail in \citet{cullen2017}.
By default we assume an escape fraction of $f_{esc}=0.0$, but we illustrate in Fig. \ref{fig_fitting_regions} how the resulting intrinsic SED is sensitive to different $f_{esc}$ assumptions.
It can be seen that at certain wavelengths (e.g. $0.25 \mu \rm{m} \lesssim \lambda \lesssim 0.35 \mu \rm{m}$) the effect of varying $f_{esc}$ can be significant.
Therefore, since the average value of $f_{esc}$ at $z\simeq3.5$ is unknown, with only a handful of individual measurements made to date \citep[e.g.][]{debarros2016,vanzella2016,shapley2016} and, furthermore, variation from galaxy to galaxy might be expected \citep[e.g.][]{paardekooper2013,paardekooper2015}, the shape of the intrinsic spectrum at these wavelengths is highly uncertain.

Another issue is the potential difference in attenuation between nebular and stellar regions.
For example, in the local Universe, and out to $z\simeq2$, studies continue to find evidence for differential attenuation, such that the nebular regions suffer a larger amount of attenuation, on average, than the stellar continuum \citep[e.g.][]{kashino2013, price2014}.
In this case Eqn. \ref{eq_dust_final} will not hold for wavelength regions with a large nebular contribution.

Indeed, properly accounting for the nebular emission when fitting the attenuation curve would require a model that incorporates variations in escape fraction, differential attenuation, and perhaps even a difference in shape of the attenuation curve between stellar and nebular regions.
We decided, therefore, to restrict our analysis to wavelengths for which the nebular contribution is low.
It is important to note, however, that the remaining fitting regions still cover both far-UV and optical wavelengths (Fig. \ref{fig_fitting_regions}), and we therefore remain sensitive to the UV to optical shape of the attenuation curve.
Our method can still distinguish, for example, between shallow Calzetti-like laws, or steep SMC-like laws, regardless of the exact details of the nebular modeling/attenuation.
We note, however, that by flagging these regions, our method is not sensitive to the presence or absence of the $2175 \mathrm{\AA}$ UV bump feature.

    \begin{figure}
        \centerline{\includegraphics[width=\columnwidth]{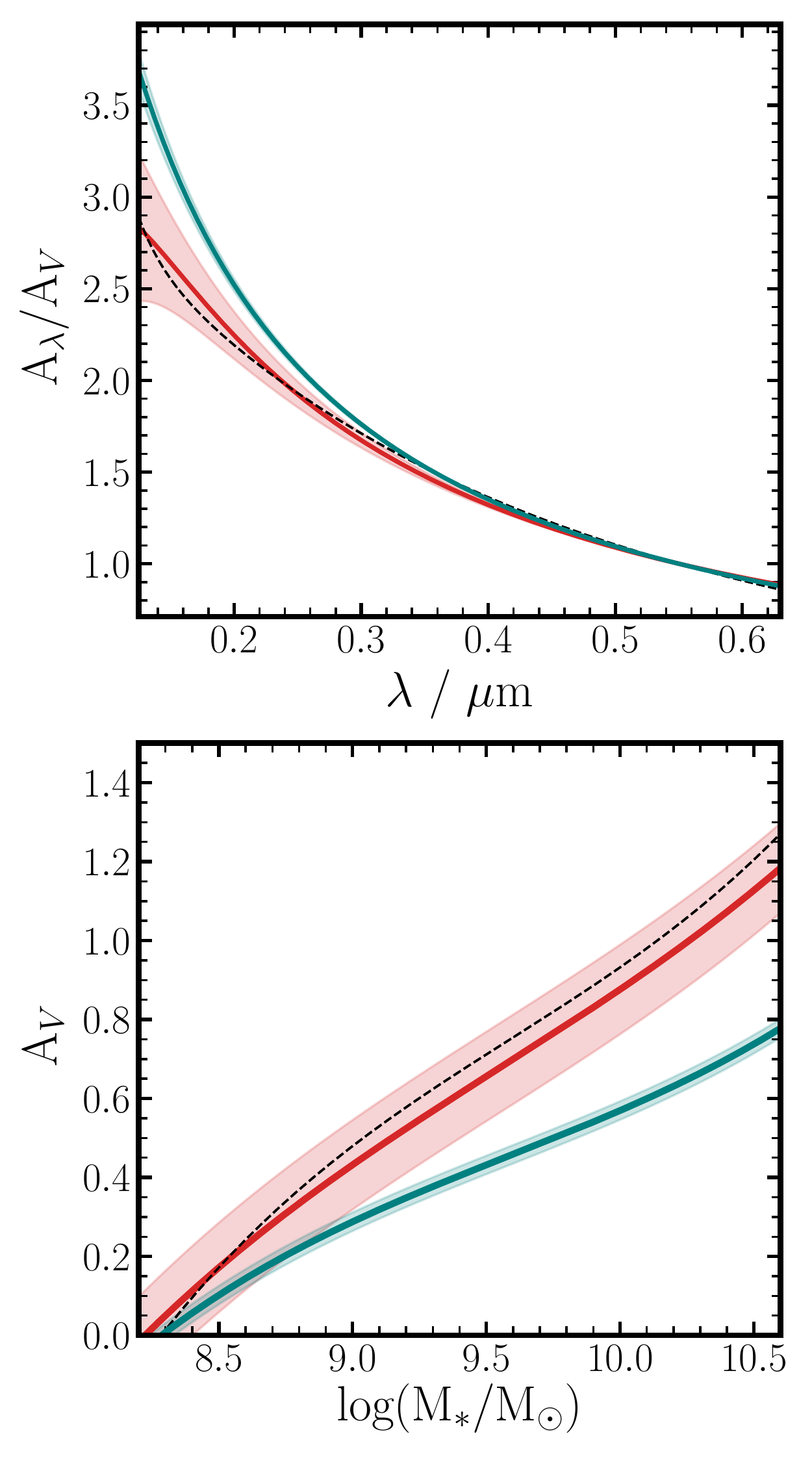}}
        \caption{Testing the accuracy of our attenuation curve fitting method with simulations. 
        The top panel shows the average attenuation curve shape ($A_{\lambda}/A_{V}$) recovered from a simulation of 1000 SEDs which have been reddened with {}a Calzetti law, drawing $A_{V}$ values from the \citet{mclure2017} $A_{V}-M_{\star}$ relation. 
        The blue line shows the $A_{\lambda}/A_{V}$ relation recovered using the full $0.12 - 0.63 \mu \rm{m}$ wavelength range in the fitting, while the red line shows the  $A_{\lambda}/A_{V}$ recovered using only the restricted wavelength regions given in Table \ref{tab_fit_windows}.
        The black dashed line shows the input $A_{\lambda}/A_{V}$ relation.
        The bottom panel shows the recovered $A_{V}-M_{\star}$ relation for both these scenarios, again compared to the input relation (black dashed line).
        In both panels, shaded regions show the $\pm 1 \sigma$ error on the recovered relations.
        }
        \label{fig_sim_attn_curve}
    \end{figure}

\subsection{Simulating the fitting method}\label{sec_attn_curve_sims}

To assess how accurately the shape and normalization of an input attenuation law could be recovered using this fitting method we performed a simple simulation.
Using the FiBY-BPASSv2-100bin SED as the intrinsic template, we constructed a sample of artificially reddened SEDs.
First, we generated 1000 values of $M_{\star}$ (equally spaced between $8.2 \leq \rm{log(M_{\star}/M}_{\odot}) \leq 10.6$) and assigned a value of $A_{V}$ using in the $A_{V}-M_{\star}$ relation for a Calzetti law from \citet{mclure2017}.
This relation is shown as the dashed black curve in the bottom panel of Fig. \ref{fig_sim_attn_curve}.
For each value of $M_{\star}$ we constructed an attenuated SED in the following way. 
The purely stellar component of the FiBY-BPASSv2-100bin template was attenuated using a Calzetti law with its corresponding $A_{V}$ value, while the purely nebular component was also attenuated with a Calzetti law, however this time assuming $A_{V,neb}=2 \times A_{V}$.
The final simulated SED was taken as the sum of the two components, allowing us to mimic the effect of differential reddening in stellar continuum and nebular regions.
Finally, we perturbed the flux values in the SED assuming a $20\%$ error in each pixel to mimic the combined uncertainty in the photometry and choice of intrinsic SED shape.

After generating 1000 artificial SEDs, the next step was to recover the attenuation curve of the stellar continuum using the method described above.
We first performed the fitting without masking any wavelength regions, the results of which are shown by the blue curves in Fig. \ref{fig_sim_attn_curve}.
We found that the underlying Calzetti law attenuating the stellar continuum could not be recovered in this case because the fit is biased by pixels with a strong nebular contribution, and/or pixels with strong stellar absorption/emission features as described above.
The recovered shape of the resulting attenuation curve ($A_{\lambda}/A_{V}$) is steeper than a Calzetti law, and the $A_{V}-M_{\star}$ relation is heavily biased to low $A_{V}$ values (by up to $\simeq 0.6$ dex at the largest values of $M_{\star}$).

In contrast, by using the wavelength regions given in Table \ref{tab_fit_windows} it is possible to recover the input relations (red curves in Fig. \ref{fig_sim_attn_curve}).
There are still small biases in the recovered $A_{V}$ which increase with $M_{\star}$, however this bias does not strongly affect the recovered attenuation curve shape ($A_{\lambda}/A_{V}$).
We note that these simulation results hold for all other intrinsic templates, and are also independent of the assumed attenuation law (i.e. similar results are found using an input SMC-like curve).
Finally, we found that although it is common to parameterize the attenuation curve using a third-order polynomial \citep[e.g.][]{calzetti2000,reddy2015}, the parameterization of $A_{\lambda}$ given by Equation \ref{eq_neta} works better at recovering $A_{V}$ and $A_{\lambda}/A_{V}$ given the restricted wavelength range we used in our fitting.
As can be seen, this parameterisation nevertheless results in a very similar attenuation curve shape to the actual Calzetti law.

In reality there will of course be a scatter in the value of $A_{V}$ at a given $M_{\star}$, in the ratio of nebular to stellar attenuation, and even in the shape of the attenuation curve which we have not captured here; however, we argue that this simple simulation illustrates how the underlying shape and normalization of the attenuation curve, as well as the $A_{V}-M_{\star}$ relation, can be robustly derived using the method outlined in Section \ref{sec_attn_curve_derivation}.

    \begin{figure*}
        \centerline{\includegraphics[width=6.7in]{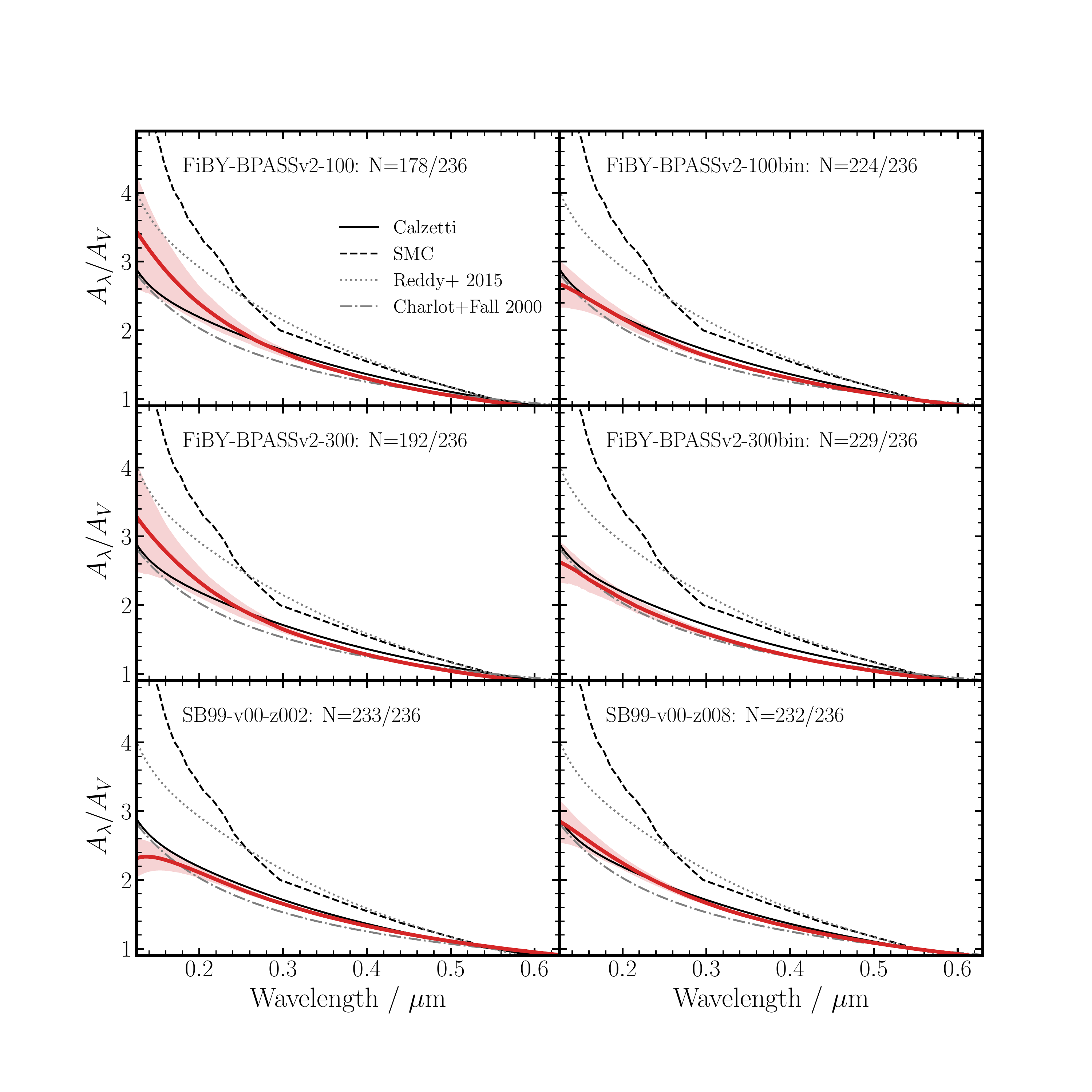}}
        \caption{Each panel shows the mean best-fitting attenuation curve shape ($A_{\lambda}/A_{V}$) corresponding to the six single-star intrinsic SED templates (red solid lines).
        The template corresponding to each line is indicated in each panel, along with the number of galaxies in the sample which are individually fitted by that template (see text for discussion).
        The shaded regions around each line show the $\pm 1 \sigma$ scatter about the mean relation.
        In each panel the black lines show four literature attenuation curves as indicated in the top left-hand panel.}
        \label{fig_full_sample_attn_curve}
    \end{figure*}

\section{Results: the UV $-$ optical attenuation curve}\label{sec_results}

In the section we present the shape and normalization of the attenuation curve at $z\simeq3.5$, derived using the method described above.

\subsection{Average shape of the attenuation curve}

We first investigated the average shape of the attenuation curve for the $236$ galaxies in our VANDELS sample.
In Fig. \ref{fig_full_sample_attn_curve} we show the average attenuation curve shapes ($A_{\lambda}/A_{V}$) corresponding to each of the six intrinsic template SEDs listed in Table \ref{tab_templates}.
For each template, we individually fitted every galaxy in the sample (an example of which is given in Fig. \ref{fig_example_fitting}), and in Fig. \ref{fig_full_sample_attn_curve} we plot the unweighted mean of all the individual curves along with the $\pm 1 \sigma$ scatter.

For each of the intrinsic templates it was not possible to find an individual solution for every galaxy in the sample with that template.
This occurred when the UV/optical ratio was larger in the observed SED than in the intrinsic SED (i.e. the observed SED was bluer), implying a negative attenuation was required to map the intrinsic to observed shape.
In these cases, either the assumed intrinsic shape, or the observed shape derived from the photometry, is clearly incorrect.
For all these cases, we assumed $A_V=0$ and excluded these galaxies when averaging. 
The degree of success of each template at fitting individual galaxies varied, and the number of successful fits is indicated in Fig. \ref{fig_full_sample_attn_curve}.
It can be seen that for half of the templates the number of unsuccessful fits was a negligible fraction of the sample ($\lesssim 5 \%$).

It is clear from Fig. \ref{fig_full_sample_attn_curve} that the majority of attenuation curves are shallow and Calzetti-like in shape.
Steeper curves are suggested from the FiBY-BPASSv2 single-star models (100, 300) however these are still not, on average, as steep as the SMC extinction law or even the \citet{reddy2015} curve.
Furthermore, the single star models have the lowest success rates for individual fits ($\approx 75 - 80\%$) indicating that these intrinsic SED shapes, which have the lowest UV/optical ratios (Fig. \ref{fig_intrinsic_templates}), are perhaps not accurate representations of the intrinsic $z\simeq3.5$ population.
Both the Starburst99 templates and FiBY-BPASSv2 binary-star models have much higher individual success rates ($\approx 90 - 99 \%$) and clearly favour a shallower curve similar to the \citet{calzetti2000} law, or the \citet{charlot2000} model.
Overall, the simple constant star-formation rate Starburst99 and the FiBY-BPASSv2-300bin templates have the highest success rate for individual fits.

Unfortunately, this analysis alone cannot distinguish which of the intrinsic templates is most likely to be correct one.
Nevertheless, it does demonstrate that, irrespective of intrinsic template choice, assumed star-formation history, or metallicity, the average shape of the attenuation curve at $z\simeq3.5$ is consistent with a grey Calzetti-like law within $\pm 1 \sigma$.
In other words, the ratio of UV (evaluated at $\lambda=0.12\mu$m) to optical ($\lambda=0.55\mu$m) attenuation is consistently within the range $A_{UV}/A_{V} \simeq 2.5 - 3.5$.
Extremely steep attenuation curves, similar in shape to the SMC extinction law ($A_{UV}/A_{V} \simeq 6.5$), are strongly disfavoured.
Finally, it is interesting to note that our results are generally incompatible with the direct measurement of the attenuation curve at $z\simeq 2$ by \citet{reddy2015} which has $A_{UV}/A_{V} \simeq 4.0$ (dotted line in Fig. \ref{fig_full_sample_attn_curve}).
While the shape of the \citet{reddy2015} curve in the UV ($\lesssim 0.26 \mu \rm{m}$) is consistent with Calzetti \citep[see e.g.][]{reddy2015,cullen2017}, the full UV to optical shape is significantly steeper.
The reason for this discrepancy is not immediately obvious, although it is plausibly related to the difference between the two methodologies. \citet{reddy2015} do not assume an intrinsic SED shape but instead rank galaxies by increasing amount dust attenuation (using the Balmer decrement as a proxy), and use ratios of stacked SEDs, as a function of absolute attenuation, to derive the shape of the attenuation curve.
As we discuss in detail below, growing evidence for a relation between stellar mass and attenuation opens up the possibility that a similar analysis to \citet{reddy2015} can be carried out using stellar mass, rather than Balmer decrement, as a proxy for absolute attenuation.
In future, utilizing the full VANDELS dataset, it should be possible to directly compare the two different approaches.

\subsection{The $\mathbf{A_V - M_{\star}}$ relation}\label{sec_av_mass}

    \begin{figure*}
        \centerline{\includegraphics[width=5.in]{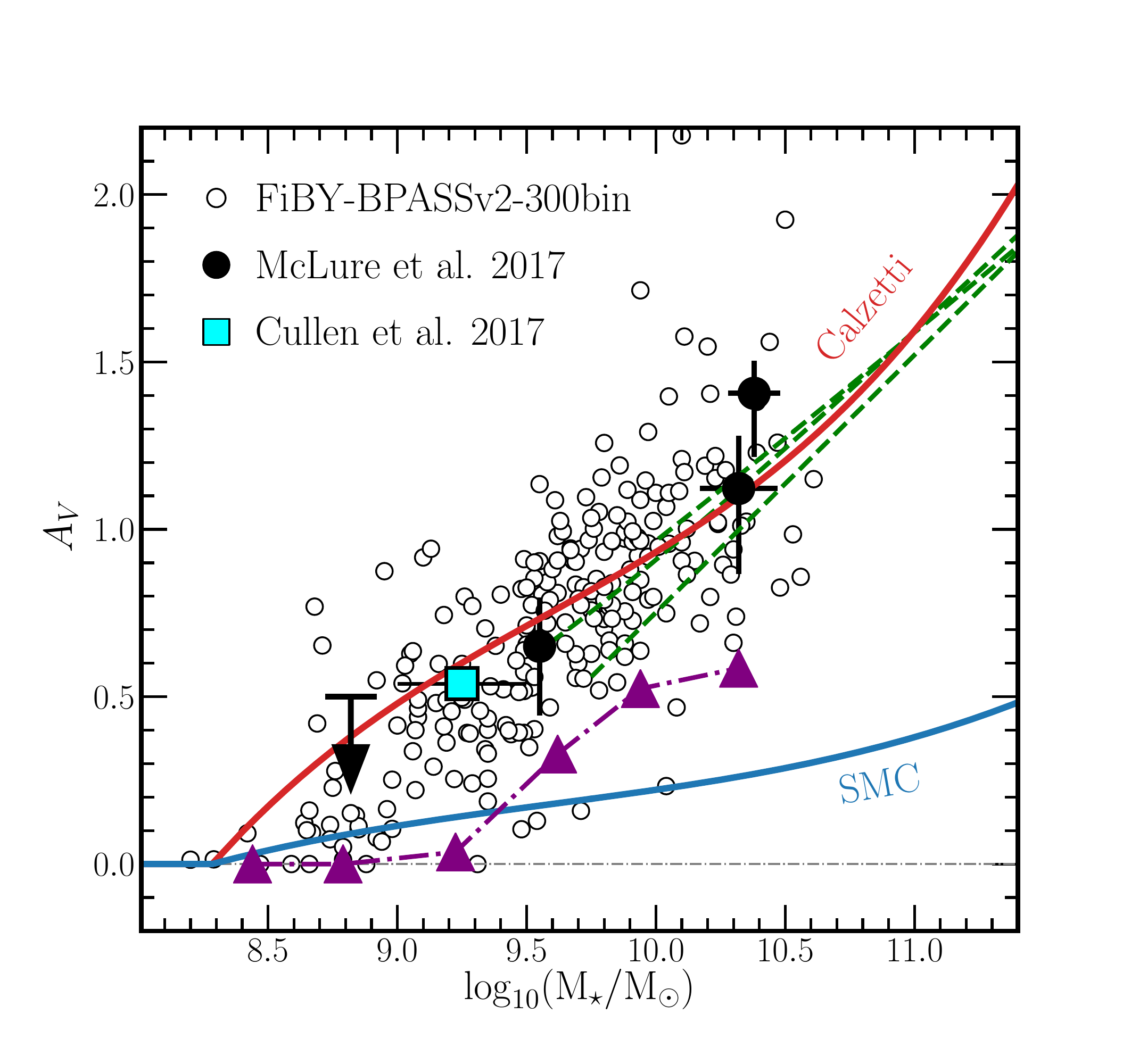}}
        \caption{$A_V$ as a function of stellar mass for the FiBY-BPASSv2-300bin template (open circular symbols).
        The red and blue lines show the empirical $A_V$ versus stellar mass relations derived by \citet{mclure2017} at $2<z<3$ from an analysis of $8407$ galaxies at $2 < z < 3$ across the stellar mass range $8.25 \leq \mathrm{log}(M_{\star}/M_{\odot}) \leq 11.50$.
        The red line assumes a \citet{calzetti2000} attenuation curve and the blue line assumes an SMC-like attenuation curve (see text for details).
        The large black circles with error bars show measurements from \citet{mclure2017} using stacked ALMA data in the \emph{Hubble} \emph{Ultra} \emph{Deep} \emph{Field}.
        The three green dashed lines are a selection of curves compiled by \citet{mclure2017} from other literature studies of galaxies between $1<z<3$.
        The cyan square is the average $A_V$ between $9.0 \leq \mathrm{log}(M_{\star}/M_{\odot}) \leq 9.5$ taken from the $A_{V}-M_{\star}$ relation derived by \citet{cullen2017} at $z\simeq5$.
       	The purple triangles show the running median $A_{V}-M_{\star}$ relation for the FiBY-BPASSv2-100 template.}
        \label{fig_av_mass}
    \end{figure*}

One way to further discriminate between these different models is to investigate the predicted absolute attenuation as a function of stellar mass, which we discuss below.
A major strength of our fitting method is that it returns an accurate estimate of $A_V$ for each galaxy (Fig. \ref{fig_sim_attn_curve}), which can be used to investigate the relationship between absolute attenuation and stellar mass ($M_{\star}$).
In Fig. \ref{fig_av_mass} we show the $A_{V}-M_{\star}$ relation returned from our fitting method for the FiBY-BPASSv2-300bin templates.
We compare this result to recent measurements of $A_{V}-M_{\star}$ at $2<z<3$ by \citet{mclure2017}, based on a deep Atacama Larges Millimeter Array (ALMA) continuum mosaic of the \emph{Hubble} \emph{Ultra} \emph{Deep} \emph{Field} \citep[HUDF;][]{dunlop2017}.
The \citet{mclure2017} data points shown in Fig. \ref{fig_av_mass} are inferred from measuring the infrared excess ($\rm{IRX} \equiv \rm{L_{IR}}/\rm{L_{UV}}$) using a stacking analysis.
Crucially, these measurements provide an independent comparison sample based on a direct detection of dust emission in the infrared.

In Fig. \ref{fig_av_mass} we also show two empirical $A_V - M_{\star}$ relations, one corresponding to the assumption of a \citet{calzetti2000} law, and the other to an SMC-like attenuation curve.
These are derived by \citet{mclure2017} from an analysis of the $M_{\star} -$ $\beta$ relation for a sample of $8407$ galaxies at $2 < z < 3$ across the stellar mass range $8.25 \leq \mathrm{log}(M_{\star}/M_{\odot}) \leq 11.50$.
We refer the reader to their paper for a full description.
It is sufficient here to note that the functional form of these relations is dependent on the shape of attenuation curve and the assumed intrinsic UV continuum slope.
\citet{mclure2017} use $\beta_{\rm{int}}=-2.3$ at $z\simeq2.5$, which is consistent with the UV continuum slopes of all of our intrinsic template SEDs ($-2.41 \leq \beta_{\rm{int}} \leq -2.31$).
As is evident from Fig. \ref{fig_av_mass}, \citet{mclure2017} found that their data follow an $A_V - M_{\star}$ consistent with a grey Calzetti-like attenuation law.
These data and empirical relations provide an interesting comparison for the template-dependent $A_V - M_{\star}$ relations which result from our analysis. 

It can be seen from Fig. \ref{fig_av_mass} that the $A_V - M_{\star}$ prediction for the FiBY-BPASSv2-300bin template is in excellent agreement with both the \citet{mclure2017} data, and the empirical relation for a Calzetti dust law at $\mathrm{log}(M_{\star}/M_{\odot}) \gtrsim 9.5$.
This may seem unsurprising given the fact that the derived attenuation curve shape for this template is so similar to the Calzetti law (Fig. \ref{fig_full_sample_attn_curve}).
However, it is important to note that this level of consistency is not necessarily guaranteed. 
Firstly, the \citet{mclure2017} data points are derived from an independent sample of galaxies, at a slightly lower redshift, using a completely different method to the one presented here.
Secondly, the empirical relation derived by \cite{mclure2017} depends only on the assumption that the shape of the UV continuum ($0.12 - 0.25 \mu \mathrm{m}$) is constant over the full range of stellar mass, while our method is based on the premise that the intrinsic UV $-$ optical ($0.12 - 0.63 \mu \mathrm{m}$) shape is unchanging.
Therefore, the excellent agreement between these separate analyses is both an argument in favour of (i) the Calzetti law being representative of the average attenuation curve at $2 < z < 4$ (at least for $\mathrm{log}(M_{\star}/M_{\odot}) \gtrsim 9.5)$, and (ii) our fundamental assumption being correct (i.e. that the intrinsic UV $-$ optical SED shape for star-forming galaxies at these redshifts is relatively constant).
Finally, despite only showing the FiBY-BPASSv2-300bin template in Fig. \ref{fig_av_mass}, we note that both the FiBY-BPASSv2-100bin and S99-v00-z008 templates give very similar results. 
In contrast, the S99-v00-z002 template and both the FiBY-BPASSv2 single-templates are incompatible with the \citet{mclure2017} data, as illustrated by the running median $A_V - M_*$ relation for the FiBY-BPASSv2-100 template in Fig. \ref{fig_av_mass}.

\subsubsection{Redshift dependence of $A_V - M_{\star}$}

In Fig. \ref{fig_av_mass} we also show three $A_V - M_*$ relations compiled from various literature sources and presented in \citet{mclure2017}.
The green curves are taken from \citet{heinis2013}, \citet{pannella2014} and \citet{alvarez_marquez2016} and are based on IR detections of dust emission from star-forming galaxies at $1 < z < 3$.
\citet{mclure2017} also show how these relations are in remarkably good agreement with attenuation inferred for star-forming galaxies drawn from the SDSS presented by \citet{garn2010}.
The consistency between our data and these studies is striking given the variety of different methods involved. 

Finally, the square cyan data point shows the average $A_V$ between $9.00 \leq \mathrm{log}(M_{\star}/M_{\odot}) \leq 9.5$ taken from the $A_{V}-M_{\star}$ relation inferred at $z\simeq5$ by \citet{cullen2017}. 
In \citet{cullen2017}, we derived $A_{V}-M_{\star}$ by comparing the observed $z\simeq5$ luminosity function and color-magnitude relation with predictions from the FiBY simulation.
This stellar-mass range is chosen to cover the masses across which both this study and \citet{cullen2017} benefit from robust statistics.
Again, the consistency is remarkable, given the differences in the two methods.
Interestingly, combining all of these literature results with our data suggests both that stellar mass is a good proxy for attenuation, and that the relationship between mass and attenuation, for normal star-forming galaxies with $\rm{log(M_{\star} / M_{\odot})} \gtrsim 9.5$, does not evolve between $z=0$ to $z\simeq5$.

In summary, the $A_V - M_{\star}$ relations predicted by the FiBY-BPASSv2 binary star templates (100bin, 300bin), and S99-v00-z008 template, are fully consistent with observed data at $1<z<3$.
These results support the view that star-forming galaxies at these redshifts are reddened, on average, by a shallow attenuation law similar to the Calzetti starburst law.
The consistency of our results with data from previous studies at different redshifts, using different techniques, is also remarkable. 
Current data appear to support a scenario in which the $A_V - M_{\star}$ relation for star-forming galaxies does not evolve significantly from the local Universe out to $z\simeq5$. 
Deeper infrared data are needed in order to make robust statements at $\rm{log(M_{\star} / M_{\odot})} < 9.5$.

    \begin{figure}
        \centerline{\includegraphics[width=\columnwidth]{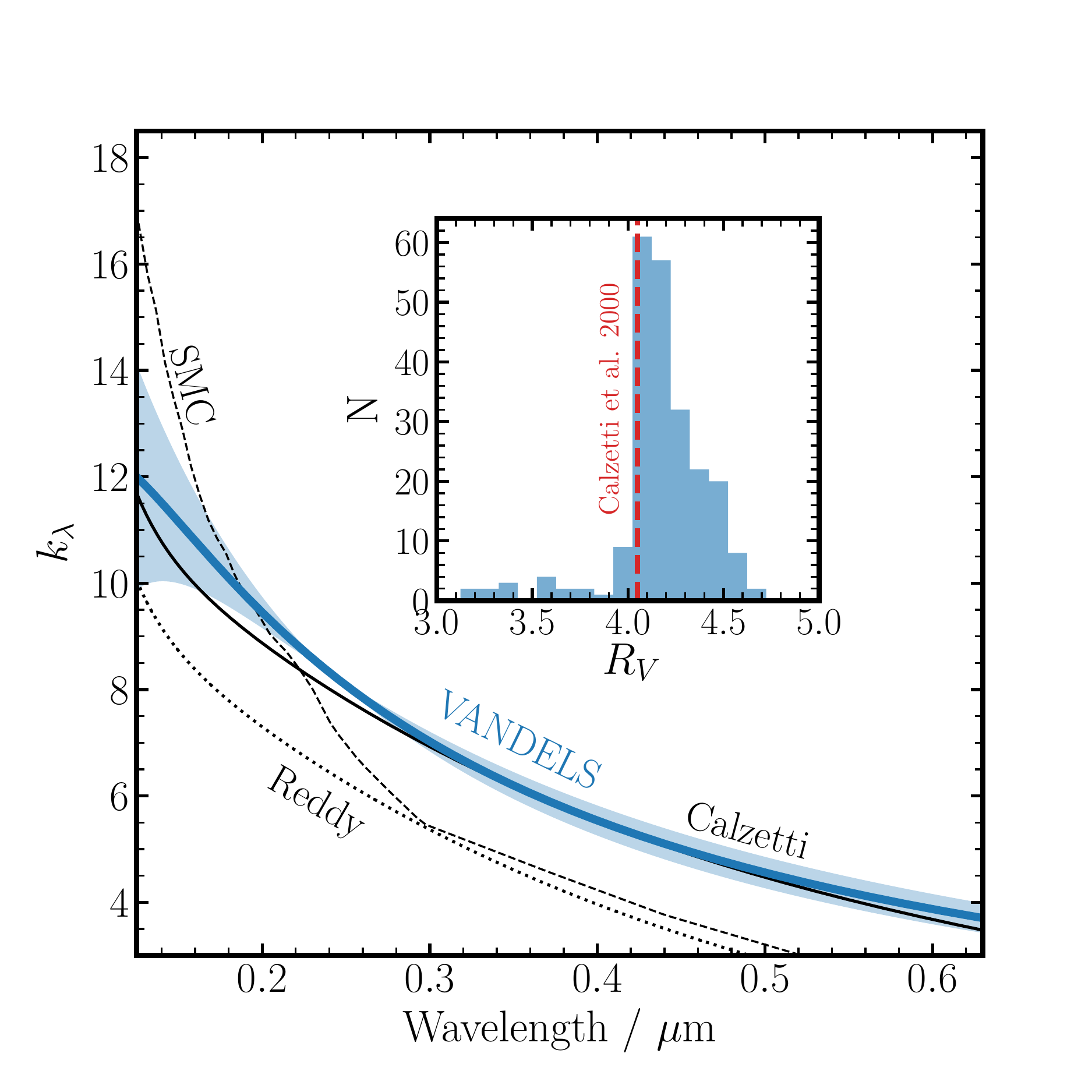}}
        \caption{The total-to-selective attenuation curve $k_{\lambda}$ for our fiducial intrinsic SED template (FiBY-BPASSv2-300bin).
        Individual $k_{\lambda}$ curves were derived using Equation \ref{eq_klam} for each of the 229 galaxies which had an attenuation curve solution.
        The blue line shows the average across all individual galaxies, with the shaded region showing the $\pm 1 \sigma$ scatter.
        A selection of $k_{\lambda}$ curves from the literature are also shown and labeled.
        The inset panel shows the distribution of $R_V$ values for all 229 galaxies obtained using Equation \ref{eq_rv}.
        The average across all galaxies is $R_V=4.18\pm0.29$, which is in good agreement with the original determination for a Calzetti law of $R_V=4.05\pm0.80$ \citep[][indicated by the dashed orange line]{calzetti2000}.}
        \label{fig_klam_rv}
    \end{figure}

\subsection{Parameterization of the attenuation curve}

Taking FiBY-BPASSv2-300bin as our fiducial intrinsic SED template we find the following parameterisations of the average attenuation law at $z \simeq 3.5$:
\begin{equation}\label{eq_final_alam_av}
\frac{A_{\lambda}}{A_V} = \frac{0.587}{\lambda} - \frac{0.020}{\lambda^2},
\end{equation}
valid in the wavelength range $0.12 < \lambda < 0.63 \ \mu \rm{m}$, with uncertainties on the parameters of around $2 \%$.

A more common parameterisation of dust attenuation/extinction curves is via the total-to-selective attenuation ratio $R_V$ and the total-to-selective attenuation curve $k_{\lambda}$.
The equations that define these quantities are given by Equations \ref{eq_rv} and \ref{eq_klam}.
In Fig. \ref{fig_klam_rv} we show the average $k_{\lambda}$ for the FiBY-BPASSv2-300bin template which can be parameterized as:
\begin{equation}\label{eq_final_klam}
k_{\lambda} = \frac{2.454}{\lambda} - \frac{0.084}{\lambda^2},
\end{equation}
and $R_V=4.18\pm0.29$. These two separate parameterizations are simply related by:
\begin{equation}
\frac{A_{\lambda}}{A_V} = \frac{k_{\lambda}}{R_V},
\end{equation}
which can be readily seen by comparing Equations \ref{eq_final_alam_av} and \ref{eq_final_klam}, using the derived value of $R_V$.
Finally, another commonly used parameter, the color excess ($E(B-V)=A_B-A_V$), is related to these various quantities via:
\begin{equation}
E(B-V) = \frac{A_{\lambda}}{k_{\lambda}} = \frac{A_V}{R_V}.
\end{equation}
As can be seen from Fig. \ref{fig_klam_rv}, the results are fully consistent with the Calzetti attenuation law. 
We find an average $R_V$ comparable with the \citet{calzetti2000} estimate of $R_V=4.05\pm0.80$.

\subsection{Attenuation curve at low masses}\label{sec_attn_low_mass}

There is a suggestion from Fig. \ref{fig_av_mass}, both from our data and the measurements of \citet{mclure2017}, that the average attenuation curve may be steeper than the Calzetti law at the lowest masses, particularly below $\mathrm{log}(M_{\star}/M_{\odot}) \simeq 9.0$.
The idea that the attenuation curve could be mass dependent has been suggested by some studies in the literature, which argue that the attenuation curve steepens as the optical depth decreases \citep[e.g.][]{seon2016,leja2017}, which could be the result of a mass-dependent change in the dust geometry in galaxies \citep[e.g.][]{paardekooper2015}.
On the other hand, the opposite trend has been suggested in other papers \citep[e.g.][]{zeimann2015_dust}.

\begin{figure}
        \centerline{\includegraphics[width=\columnwidth]{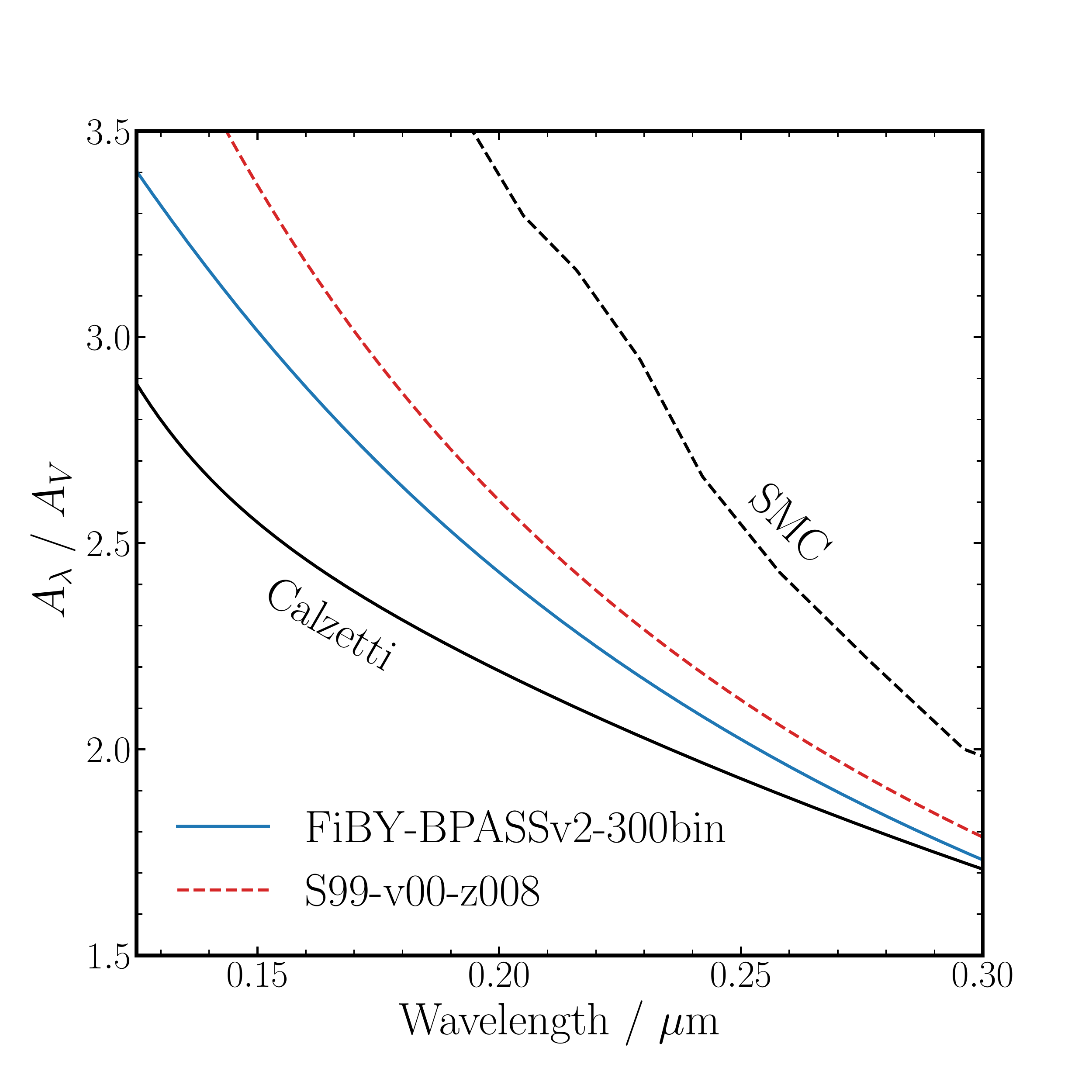}}
        \caption{The average attenuation curve shape ($A_{\lambda}/A_{V}$) for galaxies with $\mathrm{log}(M_{\star}/M_{\odot}) < 9.0$.
        The solid blue line shows our fiducial FiBY-BPASSv2-300bin template and the dashed red line shows the S990-v00-z008 template.
        For comparison we show the Calzetti law (solid black line) and the SMC extinction law (dashed black line).
        It can be seen that we find some tentative evidence for a steepening of the attenuation curve at the lowest masses, though still no evidence for an attenuation law as steep as the SMC extinction curve.}
        \label{fig_alam_mass}
    \end{figure}

In Fig. \ref{fig_alam_mass} we show the average attenuation curve shapes for the $33$ galaxies in our sample with $\mathrm{log}(M_{\star}/M_{\odot}) < 9.0$.
To improve the accuracy of this fitting we constructed stacked intrinsic spectra from FiBY with masses in the same mass range, however the results are similar using the original stacks.
It can be seen that we find some evidence for a slight steepening using our fiducial FiBY-BPASSv2-300bin template. 
The steepening is more pronounced for the S99-v00-z008 template.
However, although we don't show the $\pm 1 \sigma$ scatter of the curves in Fig. \ref{fig_alam_mass} for clarity, both remain consistent with the Calzetti attenuation law within $1 \sigma$, and rule out and SMC-like curve at $> 1 \sigma$.
Therefore, though we find some tentative evidence for a steepening of the attenuation curve at $\mathrm{log}(M_{\star}/M_{\odot}) < 9.0$, a larger sample of galaxies at these masses is needed to confirm this result.
Moreover, we still find no evidence to support an average attenuation curve as steep as the SMC extinction law, even at these lower masses.

\subsection{Evidence for a $2175\rm{\AA}$ UV bump?}

    \begin{figure}
        \centerline{\includegraphics[width=\columnwidth]{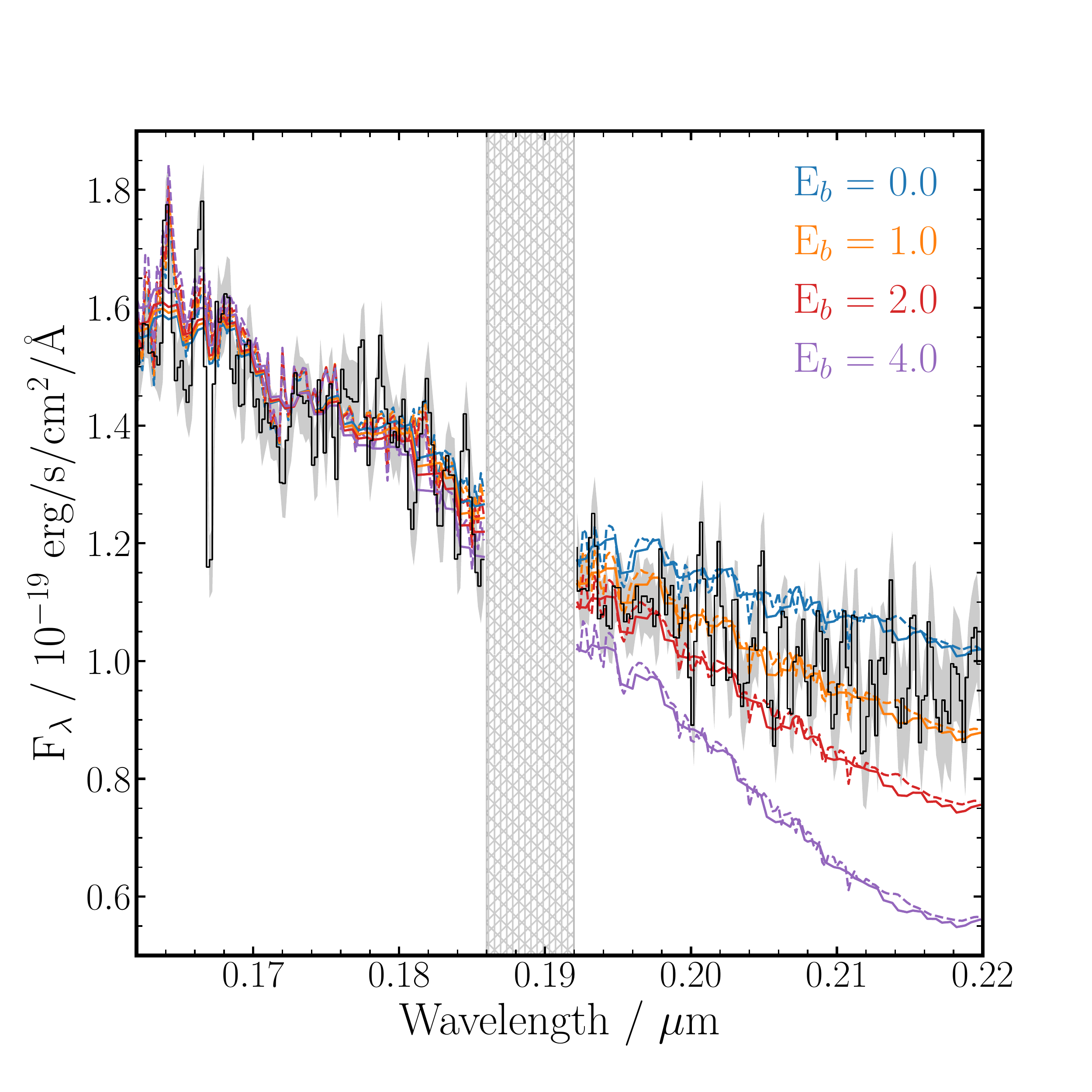}}
        \caption{Investigating evidence for excess attenuation at $2175\rm{\AA}$ from a composite of 122 VANDELS spectra at $3.0 \leq z \leq 3.5$ ($\langle z \rangle = 3.288$).
        The grey spectrum is the composite VANDELS spectrum with the grey shaded region representing the $\pm 1\sigma$ error. 
        The hatched region around $\lambda \sim 0.19 \mu \rm{m}$ masks the \ciii \ emission line.
        The solid and dashed coloured curves show, respectively, the FiBY-BPASSv2-300bin and S99-v00-z008 templates reddened using the \citet{noll2009} attenuation curve assuming four values of the $2175\rm{\AA}$ bump strength (E$_b$; see text for details).
        For each value of E$_b$ the templates have been attenuated with $A_V=0.8$ assuming the Calzetti $A_V - M_*$ relation at the median mass of the galaxies contributing to the stack ($\mathrm{log}(M_{\star}/M_{\odot}) = 9.71$; see Fig. \ref{fig_av_mass}).}
        \label{fig_uv_bump}
    \end{figure}

Finally, we investigated whether there was any evidence for excess attenuation in our sample caused by the $2175\rm{\AA}$ absorption feature \citep[often referred to as the UV bump;][]{stecher1965}.
The $2175\rm{\AA}$ absorption feature is observed in the Milky Way (MW) and Large Magellanic Cloud (LMC) extinction curves, and has recently been shown to be present in the extinction curves along the majority of sight-lines to the SMC \citep{hagen2017}.
Intriguingly, the feature is not present in the Calzetti starburst attenuation law, although a number of other studies, both in the local Universe and up to $z\sim2$, have found evidence for its presence \citep[e.g.][]{noll2009,conroy2010,wild2011,buat2011,kriek2013,battisti2017b}.
The absence/presence of a UV bump, and its dependence on other physical parameters, is a potentially useful constraint on the grain properties and physical conditions within star-forming galaxies \citep[e.g.][]{seon2016}.

As discussed in Section \ref{sec_attn_curve_derivation}, our method for deriving the shape of the UV to optical attenuation curve explicitly excluded the UV bump wavelength region.
However, the availability of deep rest-frame ultraviolet spectra for the galaxies in our sample provides an alternative route to determine whether the average attenuation curve at $z>3$ contains a $2175\rm{\AA}$ UV bump feature.
The feature is not sampled by the VIMOS spectra for galaxies at $z>3.5$, therefore we could only utilize the 122 galaxies at $3.0 < z < 3.5$ for this analysis (with $\langle z \rangle = 3.288$).
A stacked composite UV spectrum was formed by shifting each spectrum to the rest-frame, re-sampling onto a common wavelength grid and taking the median at each dispersion point after rejecting $5\sigma$ outliers.
To correct for redshift differences, the flux of each individual spectrum was scaled to the flux that would be observed at the mean redshift of the sample.
The $1\sigma$ error spectrum was estimated by bootstrap re-sampling the flux values at each dispersion point.

The composite spectrum was compared to the FiBY-BPASSv2-300bin and S99-v00-z008 templates, which were attenuated using the \citet{noll2009} parameterization of the Calzetti law:
\begin{equation}\label{eq_noll}
A(\lambda)=\frac{A_V}{4.05}(k(\lambda)_{\rm{Calzetti}} + D(\lambda)),
\end{equation}
where $D(\lambda)$ is the Drude profile parameterization of the UV bump, defined as:
\begin{equation}
D(\lambda)=\frac{E_b(\lambda \Delta \lambda)^2}{(\lambda^2-\lambda_0^2) + (\lambda \Delta \lambda)^2}.
\end{equation}
For the central wavelength we adopt $\lambda_0=2175 \rm{\AA}$ and for the full width half maximum (FWHM) we use $\Delta \lambda=350 \rm{\AA}$ \citep{noll2009}.
The parameter $E_b$ determines the strength of the feature.
We assumed $A_V=0.8$ by taking the median mass of the galaxies contributing to the stack ($\mathrm{log}(M_{\star}/M_{\odot}) = 9.71$) and assuming the Calzetti $A_{V}-M_{\star}$ relation from Fig. \ref{fig_av_mass}.

The two intrinsic templates were attenuated using Equation \ref{eq_noll} for various values of $E_b$ and compared to the composite spectrum after normalizing to the median flux in the wavelength range $0.17 < \lambda < 0.18\mu \rm{m}$ (Fig. \ref{fig_uv_bump}).
The data robustly rule out the presence of a strong UV bump with $E_b > 1.0$.
Interestingly, assuming a weak $2175\rm{\AA}$ UV bump feature with $E_b \simeq 1$ restults in a better match to the observed composite spectrum than a pure Calzetti law ($E_b=0$).
This result is consistent with the K-band selected sample of galaxies at $0.5 < z < 2.5$ from \citet{kriek2013}, who find that the strength of the UV bump is correlated with the steepness of the attenuation curve, and that for Calzetti-like curves $E_b = 0.85 \pm 0.09$.
However, given the uncertainties on the composite spectrum, we consider this only marginal evidence for a UV bump. 
Nevertheless, the comparison does constrain the average strength of the UV bump to be, on average, relatively weak at $z \simeq 3$ ($E_b \lesssim 1$).
For comparison, the UV bump in the Milky Way extinction curve has $E_b \simeq 4$.
In a future work, with the full VANDELS sample, we will present a thorough analysis of the presence/absence of a UV bump at $2.5 < z < 3.5$, and its dependence on various galaxy properties.

\section{Conclusions}\label{sec_conclusions}

We have presented an investigation into the attenuation curve of galaxies at $z=3-4$ using an initial sample of $236$ star-forming galaxies from the VANDELS survey.
By comparing the observed shape of the galaxy SEDs to a set of intrinsic SED templates we have been able to derive the shape and normalization of the attenuation curve based on the assumption that all galaxies across the mass range $8.2 < \rm{log(M_{\star} / M_{\odot})} < 10.6$ have the same underlying intrinsic SED shape.
Using this method we investigated the average attenuation curve at $z\simeq3.5$ and its mass dependence, as well as the relation between absolute attenuation and stellar mass.
Our main findings can be summarized as follows:

\begin{enumerate}

\item From an analysis of the star-formation histories and metallicities of a sample of $628$ simulated galaxies extracted from the FiBY simulation at $z=4$, we argue that the intrinsic shape of the UV $-$ optical SED of star-forming galaxies at these redshifts is approximately constant across the full stellar-mass range of our observed sample.
We construct a set of six intrinsic SED templates, both using physically-motivated star-formation histories and metallicities from FiBY, as well as simple Starburst99 constant star-formation rate models.

\item Based on the assumption of a constant intrinsic SED shape, we outline a method for fitting the shape and normalization of the attenuation curve for any individual observed galaxy SED. 
We demonstrate the feasibility of this method with a simple simulation.
Applying this method to the full sample we find that the average attenuation curve shapes derived from each of the intrinsic templates are consistent with a grey Calzetti-like attenuation law within $\pm 1 \sigma$, and find no evidence for a steep SMC-like attenuation law.

\item By comparing the $A_V - M_{\star}$ relations predicted by each template to recent observations at $2<z<3$ from \cite{mclure2017}, we find that a subset of the intrinsic template sets yield results which are fully consistent with the observed data, and with the $A_V - M_{\star}$ relation predicted for a Calzetti-like attenuation law.
Again, we find no evidence for an $A_V - M_{\star}$ relation consistent with the SMC extinction law.
This result, combined with other literature data, suggest that the relationship between stellar mass and $A_V$ does not evolve over the redshift range $0<z<5$.	
Furthermore, across all these redshifts, we confirm previous results that suggest stellar mass is a good proxy for absolute attenuation.

\item Using our fiducial intrinsic template (FiBY-BPASSv2-300bin) we transform our inferred attenuation curve ($A_{\lambda}$) into the more standard total-to-selective attenuation curve formulation ($k_{\lambda}$). 
We find a total-to-selective attenuation ratio of $R_V = 4.18\pm0.29$, consistent with the original \citet{calzetti2000} value of $R_V = 4.05 \pm 0.80$.

\item We present tentative evidence for steeper attenuation curve shapes at masses $\mathrm{log}(M_{\star}/M_{\odot}) \lesssim 9.0$, in line with models which predict a dependence of the attenuation curve shape on the total optical depth.
Better statistics are required to confirm this results, or alternatively deeper ALMA continuum imaging at these masses.

\item Finally, using a stacked composite spectrum of 122 galaxies at $3.0 < z < 3.5$ we find marginal evidence for a weak $2175 \rm{\AA}$ UV bump in the average attenuation curve at $z\simeq3$, but robustly rule out the presence of a strong UV bump feature.
A more detailed examination will be possible in future using the full VANDELS spectroscopic sample.

\end{enumerate}

\section{Acknowledgments}
The authors would like to thank the anonymous referee whose comments and suggestions significantly improved the final version of this manuscript.
FC, RJM, SK and JSD acknowledge the support of the UK Science and Technology Facilities Council.
AC acknowledges the grants ASI n.I/023/12/0 `Attivit\`{a} relative alla fase B2/C per la missione Euclid' and MIUR PRIN 2015 `Cosmology and Fundamental Physics: illuminating the Dark Universe with Euclid'.
The Cosmic Dawn center is funded by the DNRF.
This research made use of Astropy, a community-developed core Python package for Astronomy \citep{astropy2013}, NumPy and SciPy \citep{oliphant2007}, Matplotlib \citep{hunter2007}, {IPython} \citep{perez2007} and NASA's Astrophysics Data System Bibliographic Services.


\bibliographystyle{mnras}                      
\bibliography{vandels_z4_dust}       

\begin{thebibliography}{}
\makeatletter
\relax
\def\mn@urlcharsother{\let\do\@makeother \do\$\do\&\do\#\do\^\do\_\do\%\do\~}
\def\mn@doi{\begingroup\mn@urlcharsother \@ifnextchar [ {\mn@doi@}
  {\mn@doi@[]}}
\def\mn@doi@[#1]#2{\def\@tempa{#1}\ifx\@tempa\@empty \href
  {http://dx.doi.org/#2} {doi:#2}\else \href {http://dx.doi.org/#2} {#1}\fi
  \endgroup}
\def\mn@eprint#1#2{\mn@eprint@#1:#2::\@nil}
\def\mn@eprint@arXiv#1{\href {http://arxiv.org/abs/#1} {{\tt arXiv:#1}}}
\def\mn@eprint@dblp#1{\href {http://dblp.uni-trier.de/rec/bibtex/#1.xml}
  {dblp:#1}}
\def\mn@eprint@#1:#2:#3:#4\@nil{\def\@tempa {#1}\def\@tempb {#2}\def\@tempc
  {#3}\ifx \@tempc \@empty \let \@tempc \@tempb \let \@tempb \@tempa \fi \ifx
  \@tempb \@empty \def\@tempb {arXiv}\fi \@ifundefined
  {mn@eprint@\@tempb}{\@tempb:\@tempc}{\expandafter \expandafter \csname
  mn@eprint@\@tempb\endcsname \expandafter{\@tempc}}}

\bibitem[\protect\citeauthoryear{{{\'A}lvarez-M{\'a}rquez}
  et~al.,}{{{\'A}lvarez-M{\'a}rquez} et~al.}{2016}]{alvarez_marquez2016}
{{\'A}lvarez-M{\'a}rquez} J.,  et~al., 2016, \mn@doi [\aap]
  {10.1051/0004-6361/201527190}, \href
  {http://adsabs.harvard.edu/abs/2016A%26A...587A.122A} {587, A122}

\bibitem[\protect\citeauthoryear{{Asplund}, {Grevesse}, {Sauval}  \&
  {Scott}}{{Asplund} et~al.}{2009}]{asplund2009}
{Asplund} M.,  {Grevesse} N.,  {Sauval} A.~J.,   {Scott} P.,  2009, \mn@doi
  [\araa] {10.1146/annurev.astro.46.060407.145222}, \href
  {http://adsabs.harvard.edu/abs/2009ARA%26A..47..481A} {47, 481}

\bibitem[\protect\citeauthoryear{{Astropy Collaboration} et~al.,}{{Astropy
  Collaboration} et~al.}{2013}]{astropy2013}
{Astropy Collaboration} et~al., 2013, \mn@doi [\aap]
  {10.1051/0004-6361/201322068}, \href
  {http://adsabs.harvard.edu/abs/2013A%26A...558A..33A} {558, A33}

\bibitem[\protect\citeauthoryear{{Battisti}, {Calzetti}  \& {Chary}}{{Battisti}
  et~al.}{2017a}]{battisti2017}
{Battisti} A.~J.,  {Calzetti} D.,   {Chary} R.-R.,  2017a, \mn@doi [\apj]
  {10.3847/1538-4357/aa6fb2}, \href
  {http://adsabs.harvard.edu/abs/2017ApJ...840..109B} {840, 109}

\bibitem[\protect\citeauthoryear{{Battisti}, {Calzetti}  \& {Chary}}{{Battisti}
  et~al.}{2017b}]{battisti2017b}
{Battisti} A.~J.,  {Calzetti} D.,   {Chary} R.-R.,  2017b, \mn@doi [\apj]
  {10.3847/1538-4357/aa9a43}, \href
  {http://adsabs.harvard.edu/abs/2017ApJ...851...90B} {851, 90}

\bibitem[\protect\citeauthoryear{{Bouwens} et~al.,}{{Bouwens}
  et~al.}{2016}]{bouwens2016}
{Bouwens} R.~J.,  et~al., 2016, \mn@doi [\apj] {10.3847/1538-4357/833/1/72},
  \href {http://adsabs.harvard.edu/abs/2016ApJ...833...72B} {833, 72}

\bibitem[\protect\citeauthoryear{Brammer, van Dokkum  \& Coppi}{Brammer
  et~al.}{2008}]{brammer2008}
Brammer G.~B.,  van Dokkum P.~G.,   Coppi P.,  2008, \apj, 686, 1503

\bibitem[\protect\citeauthoryear{{Bruzual} \& {Charlot}}{{Bruzual} \&
  {Charlot}}{2003}]{bruzual2003}
{Bruzual} G.,  {Charlot} S.,  2003, \mn@doi [\mnras]
  {10.1046/j.1365-8711.2003.06897.x}, \href
  {http://adsabs.harvard.edu/abs/2003MNRAS.344.1000B} {344, 1000}

\bibitem[\protect\citeauthoryear{{Buat} et~al.,}{{Buat}
  et~al.}{2011}]{buat2011}
{Buat} V.,  et~al., 2011, \mn@doi [\aap] {10.1051/0004-6361/201117264}, \href
  {http://adsabs.harvard.edu/abs/2011A%26A...533A..93B} {533, A93}

\bibitem[\protect\citeauthoryear{{Buat} et~al.,}{{Buat}
  et~al.}{2012}]{buat2012}
{Buat} V.,  et~al., 2012, \mn@doi [\aap] {10.1051/0004-6361/201219405}, \href
  {http://adsabs.harvard.edu/abs/2012A%26A...545A.141B} {545, A141}

\bibitem[\protect\citeauthoryear{{Calzetti}}{{Calzetti}}{2001}]{calzetti2001}
{Calzetti} D.,  2001, \mn@doi [\pasp] {10.1086/324269}, \href
  {http://adsabs.harvard.edu/abs/2001PASP..113.1449C} {113, 1449}

\bibitem[\protect\citeauthoryear{{Calzetti}, {Kinney}  \&
  {Storchi-Bergmann}}{{Calzetti} et~al.}{1994}]{calzetti1994}
{Calzetti} D.,  {Kinney} A.~L.,   {Storchi-Bergmann} T.,  1994, \mn@doi [\apj]
  {10.1086/174346}, \href {http://adsabs.harvard.edu/abs/1994ApJ...429..582C}
  {429, 582}

\bibitem[\protect\citeauthoryear{{Calzetti}, {Armus}, {Bohlin}, {Kinney},
  {Koornneef}  \& {Storchi-Bergmann}}{{Calzetti} et~al.}{2000}]{calzetti2000}
{Calzetti} D.,  {Armus} L.,  {Bohlin} R.~C.,  {Kinney} A.~L.,  {Koornneef} J.,
   {Storchi-Bergmann} T.,  2000, \mn@doi [\apj] {10.1086/308692}, \href
  {http://adsabs.harvard.edu/abs/2000ApJ...533..682C} {533, 682}

\bibitem[\protect\citeauthoryear{{Capak} et~al.,}{{Capak}
  et~al.}{2015}]{capak2015}
{Capak} P.~L.,  et~al., 2015, \mn@doi [\nat] {10.1038/nature14500}, \href
  {http://adsabs.harvard.edu/abs/2015Natur.522..455C} {522, 455}

\bibitem[\protect\citeauthoryear{{Cardelli}, {Clayton}  \& {Mathis}}{{Cardelli}
  et~al.}{1989}]{cardelli1989}
{Cardelli} J.~A.,  {Clayton} G.~C.,   {Mathis} J.~S.,  1989, \mn@doi [\apj]
  {10.1086/167900}, \href {http://adsabs.harvard.edu/abs/1989ApJ...345..245C}
  {345, 245}

\bibitem[\protect\citeauthoryear{{Carnall}, {McLure}, {Dunlop}  \&
  {Dav{\'e}}}{{Carnall} et~al.}{2017}]{carnall2017}
{Carnall} A.~C.,  {McLure} R.~J.,  {Dunlop} J.~S.,   {Dav{\'e}} R.,  2017,
  preprint, \href {http://adsabs.harvard.edu/abs/2017arXiv171204452C} {}
  (\mn@eprint {arXiv} {1712.04452})

\bibitem[\protect\citeauthoryear{{Chabrier}}{{Chabrier}}{2003}]{chabrier2003}
{Chabrier} G.,  2003, \mn@doi [\pasp] {10.1086/376392}, \href
  {http://adsabs.harvard.edu/abs/2003PASP..115..763C} {115, 763}

\bibitem[\protect\citeauthoryear{{Charlot} \& {Fall}}{{Charlot} \&
  {Fall}}{2000}]{charlot2000}
{Charlot} S.,  {Fall} S.~M.,  2000, \mn@doi [\apj] {10.1086/309250}, \href
  {http://adsabs.harvard.edu/abs/2000ApJ...539..718C} {539, 718}

\bibitem[\protect\citeauthoryear{{Conroy}, {Schiminovich}  \&
  {Blanton}}{{Conroy} et~al.}{2010}]{conroy2010}
{Conroy} C.,  {Schiminovich} D.,   {Blanton} M.~R.,  2010, \mn@doi [\apj]
  {10.1088/0004-637X/718/1/184}, \href
  {http://adsabs.harvard.edu/abs/2010ApJ...718..184C} {718, 184}

\bibitem[\protect\citeauthoryear{{Coppin} et~al.,}{{Coppin}
  et~al.}{2015}]{coppin2015}
{Coppin} K.~E.~K.,  et~al., 2015, \mn@doi [\mnras] {10.1093/mnras/stu2185},
  \href {http://adsabs.harvard.edu/abs/2015MNRAS.446.1293C} {446, 1293}

\bibitem[\protect\citeauthoryear{{Cullen}, {Cirasuolo}, {McLure}, {Dunlop}  \&
  {Bowler}}{{Cullen} et~al.}{2014}]{cullen2014}
{Cullen} F.,  {Cirasuolo} M.,  {McLure} R.~J.,  {Dunlop} J.~S.,   {Bowler}
  R.~A.~A.,  2014, \mn@doi [\mnras] {10.1093/mnras/stu443}, \href
  {http://adsabs.harvard.edu/abs/2014MNRAS.440.2300C} {440, 2300}

\bibitem[\protect\citeauthoryear{{Cullen}, {Cirasuolo}, {Kewley}, {McLure},
  {Dunlop}  \& {Bowler}}{{Cullen} et~al.}{2016}]{cullen2016}
{Cullen} F.,  {Cirasuolo} M.,  {Kewley} L.~J.,  {McLure} R.~J.,  {Dunlop}
  J.~S.,   {Bowler} R.~A.~A.,  2016, \mn@doi [\mnras] {10.1093/mnras/stw1181},
  \href {http://adsabs.harvard.edu/abs/2016MNRAS.460.3002C} {460, 3002}

\bibitem[\protect\citeauthoryear{{Cullen}, {McLure}, {Khochfar}, {Dunlop}  \&
  {Dalla Vecchia}}{{Cullen} et~al.}{2017}]{cullen2017}
{Cullen} F.,  {McLure} R.~J.,  {Khochfar} S.,  {Dunlop} J.~S.,   {Dalla
  Vecchia} C.,  2017, \mn@doi [\mnras] {10.1093/mnras/stx1451}, \href
  {http://adsabs.harvard.edu/abs/2017MNRAS.470.3006C} {470, 3006}

\bibitem[\protect\citeauthoryear{{Draine}}{{Draine}}{2003}]{draine2003}
{Draine} B.~T.,  2003, \mn@doi [\araa]
  {10.1146/annurev.astro.41.011802.094840}, \href
  {http://adsabs.harvard.edu/abs/2003ARA%26A..41..241D} {41, 241}

\bibitem[\protect\citeauthoryear{{Dunlop} et~al.,}{{Dunlop}
  et~al.}{2017}]{dunlop2017}
{Dunlop} J.~S.,  et~al., 2017, \mn@doi [\mnras] {10.1093/mnras/stw3088}, \href
  {http://adsabs.harvard.edu/abs/2017MNRAS.466..861D} {466, 861}

\bibitem[\protect\citeauthoryear{{Eldridge} \& {Stanway}}{{Eldridge} \&
  {Stanway}}{2016}]{eldridge2016}
{Eldridge} J.~J.,  {Stanway} E.~R.,  2016, \mn@doi [\mnras]
  {10.1093/mnras/stw1772}, \href
  {http://adsabs.harvard.edu/abs/2016MNRAS.462.3302E} {462, 3302}

\bibitem[\protect\citeauthoryear{{Faisst} et~al.,}{{Faisst}
  et~al.}{2017}]{faisst2017}
{Faisst} A.~L.,  et~al., 2017, \mn@doi [\apj] {10.3847/1538-4357/aa886c}, \href
  {http://adsabs.harvard.edu/abs/2017ApJ...847...21F} {847, 21}

\bibitem[\protect\citeauthoryear{{Ferland} et~al.,}{{Ferland}
  et~al.}{2017}]{ferland2017}
{Ferland} G.~J.,  et~al., 2017, \rmxaa, \href
  {http://adsabs.harvard.edu/abs/2017RMxAA..53..385F} {53, 385}

\bibitem[\protect\citeauthoryear{{Finlator}, {Oppenheimer}  \&
  {Dav{\'e}}}{{Finlator} et~al.}{2011}]{finlator2011}
{Finlator} K.,  {Oppenheimer} B.~D.,   {Dav{\'e}} R.,  2011, \mn@doi [\mnras]
  {10.1111/j.1365-2966.2010.17554.x}, \href
  {http://adsabs.harvard.edu/abs/2011MNRAS.410.1703F} {410, 1703}

\bibitem[\protect\citeauthoryear{{Fioc} \& {Rocca-Volmerange}}{{Fioc} \&
  {Rocca-Volmerange}}{1997}]{fioc1997}
{Fioc} M.,  {Rocca-Volmerange} B.,  1997, \aap, \href
  {http://adsabs.harvard.edu/abs/1997A%26A...326..950F} {326, 950}

\bibitem[\protect\citeauthoryear{{Fontana} et~al.,}{{Fontana}
  et~al.}{2014}]{fontana2014}
{Fontana} A.,  et~al., 2014, \mn@doi [\aap] {10.1051/0004-6361/201423543},
  \href {http://adsabs.harvard.edu/abs/2014A%26A...570A..11F} {570, A11}

\bibitem[\protect\citeauthoryear{{Fudamoto} et~al.,}{{Fudamoto}
  et~al.}{2017}]{fudamoto2017}
{Fudamoto} Y.,  et~al., 2017, \mn@doi [\mnras] {10.1093/mnras/stx1948}, \href
  {http://adsabs.harvard.edu/abs/2017MNRAS.472..483F} {472, 483}

\bibitem[\protect\citeauthoryear{{Fynbo} et~al.,}{{Fynbo}
  et~al.}{2014}]{fynbo2014}
{Fynbo} J.~P.~U.,  et~al., 2014, \mn@doi [\aap] {10.1051/0004-6361/201424726},
  \href {http://adsabs.harvard.edu/abs/2014A%26A...572A..12F} {572, A12}

\bibitem[\protect\citeauthoryear{{Galametz} et~al.,}{{Galametz}
  et~al.}{2013}]{galametz2013}
{Galametz} A.,  et~al., 2013, \mn@doi [\apjs] {10.1088/0067-0049/206/2/10},
  \href {http://adsabs.harvard.edu/abs/2013ApJS..206...10G} {206, 10}

\bibitem[\protect\citeauthoryear{{Garn} \& {Best}}{{Garn} \&
  {Best}}{2010}]{garn2010}
{Garn} T.,  {Best} P.~N.,  2010, \mn@doi [\mnras]
  {10.1111/j.1365-2966.2010.17321.x}, \href
  {http://adsabs.harvard.edu/abs/2010MNRAS.409..421G} {409, 421}

\bibitem[\protect\citeauthoryear{{Gordon}, {Clayton}, {Misselt}, {Landolt}  \&
  {Wolff}}{{Gordon} et~al.}{2003}]{gordon2003}
{Gordon} K.~D.,  {Clayton} G.~C.,  {Misselt} K.~A.,  {Landolt} A.~U.,   {Wolff}
  M.~J.,  2003, \mn@doi [\apj] {10.1086/376774}, \href
  {http://adsabs.harvard.edu/abs/2003ApJ...594..279G} {594, 279}

\bibitem[\protect\citeauthoryear{Grogin et~al.,}{Grogin
  et~al.}{2011}]{grogin2011}
Grogin N.~A.,  et~al., 2011, \mn@doi [\apjs] {10.1088/0067-0049/197/2/35},
  \href {http://adsabs.harvard.edu/abs/2011\apjs..197...35G} {197, 35}

\bibitem[\protect\citeauthoryear{{Guo} et~al.,}{{Guo} et~al.}{2013}]{guo2013}
{Guo} Y.,  et~al., 2013, \mn@doi [\apjs] {10.1088/0067-0049/207/2/24}, \href
  {http://adsabs.harvard.edu/abs/2013ApJS..207...24G} {207, 24}

\bibitem[\protect\citeauthoryear{{Hagen}, {Siegel}, {Hoversten}, {Gronwall},
  {Immler}  \& {Hagen}}{{Hagen} et~al.}{2017}]{hagen2017}
{Hagen} L.~M.~Z.,  {Siegel} M.~H.,  {Hoversten} E.~A.,  {Gronwall} C.,
  {Immler} S.,   {Hagen} A.,  2017, \mn@doi [\mnras] {10.1093/mnras/stw2954},
  \href {http://adsabs.harvard.edu/abs/2017MNRAS.466.4540H} {466, 4540}

\bibitem[\protect\citeauthoryear{{Hayes}, {Schaerer}, {{\"O}stlin},
  {Mas-Hesse}, {Atek}  \& {Kunth}}{{Hayes} et~al.}{2011}]{hayes2011}
{Hayes} M.,  {Schaerer} D.,  {{\"O}stlin} G.,  {Mas-Hesse} J.~M.,  {Atek} H.,
  {Kunth} D.,  2011, \mn@doi [\apj] {10.1088/0004-637X/730/1/8}, \href
  {http://adsabs.harvard.edu/abs/2011ApJ...730....8H} {730, 8}

\bibitem[\protect\citeauthoryear{{Heinis} et~al.,}{{Heinis}
  et~al.}{2013}]{heinis2013}
{Heinis} S.,  et~al., 2013, \mn@doi [\mnras] {10.1093/mnras/sts397}, \href
  {http://adsabs.harvard.edu/abs/2013MNRAS.429.1113H} {429, 1113}

\bibitem[\protect\citeauthoryear{{Heintz} et~al.,}{{Heintz}
  et~al.}{2017}]{heintz2017}
{Heintz} K.~E.,  et~al., 2017, \mn@doi [\aap] {10.1051/0004-6361/201730702},
  \href {http://adsabs.harvard.edu/abs/2017A%26A...601A..83H} {601, A83}

\bibitem[\protect\citeauthoryear{Hunter}{Hunter}{2007}]{hunter2007}
Hunter J.~D.,  2007, Computing In Science \& Engineering, 9, 90

\bibitem[\protect\citeauthoryear{Ilbert et~al.,}{Ilbert
  et~al.}{2008}]{ilbert2008}
Ilbert O.,  et~al., 2008, \apj, 690, 1236

\bibitem[\protect\citeauthoryear{{Johnson}, {Dalla Vecchia}  \&
  {Khochfar}}{{Johnson} et~al.}{2013}]{johnson2013}
{Johnson} J.~L.,  {Dalla Vecchia} C.,   {Khochfar} S.,  2013, \mn@doi [\mnras]
  {10.1093/mnras/sts011}, \href
  {http://adsabs.harvard.edu/abs/2013MNRAS.428.1857J} {428, 1857}

\bibitem[\protect\citeauthoryear{Kashino et~al.,}{Kashino
  et~al.}{2013}]{kashino2013}
Kashino D.,  et~al., 2013, \apj, 777, L8

\bibitem[\protect\citeauthoryear{{Kewley}, {Zahid}, {Geller}, {Dopita}, {Hwang}
   \& {Fabricant}}{{Kewley} et~al.}{2015}]{kewley2015}
{Kewley} L.~J.,  {Zahid} H.~J.,  {Geller} M.~J.,  {Dopita} M.~A.,  {Hwang}
  H.~S.,   {Fabricant} D.,  2015, \mn@doi [\apjl]
  {10.1088/2041-8205/812/2/L20}, \href
  {http://adsabs.harvard.edu/abs/2015ApJ...812L..20K} {812, L20}

\bibitem[\protect\citeauthoryear{Koekemoer et~al.,}{Koekemoer
  et~al.}{2011}]{koekemoer2011}
Koekemoer A.~M.,  et~al., 2011, \apjs, 197, 36

\bibitem[\protect\citeauthoryear{{Kriek} \& {Conroy}}{{Kriek} \&
  {Conroy}}{2013}]{kriek2013}
{Kriek} M.,  {Conroy} C.,  2013, \mn@doi [\apjl] {10.1088/2041-8205/775/1/L16},
  \href {http://adsabs.harvard.edu/abs/2013ApJ...775L..16K} {775, L16}

\bibitem[\protect\citeauthoryear{{Kroupa}}{{Kroupa}}{2001}]{kroupa2001}
{Kroupa} P.,  2001, \mn@doi [\mnras] {10.1046/j.1365-8711.2001.04022.x}, \href
  {http://adsabs.harvard.edu/abs/2001MNRAS.322..231K} {322, 231}

\bibitem[\protect\citeauthoryear{{Laporte} et~al.,}{{Laporte}
  et~al.}{2017}]{laporte2017}
{Laporte} N.,  et~al., 2017, \mn@doi [\apjl] {10.3847/2041-8213/aa62aa}, \href
  {http://adsabs.harvard.edu/abs/2017ApJ...837L..21L} {837, L21}

\bibitem[\protect\citeauthoryear{Leitherer, Ekstr{\"o}m, Meynet, Schaerer,
  Agienko  \& Levesque}{Leitherer et~al.}{2014}]{leitherer2014}
Leitherer C.,  Ekstr{\"o}m S.,  Meynet G.,  Schaerer D.,  Agienko K.~B.,
  Levesque E.~M.,  2014, \apjs, 212, 14

\bibitem[\protect\citeauthoryear{{Leja}, {Johnson}, {Conroy}, {van Dokkum}  \&
  {Byler}}{{Leja} et~al.}{2017}]{leja2017}
{Leja} J.,  {Johnson} B.~D.,  {Conroy} C.,  {van Dokkum} P.~G.,   {Byler} N.,
  2017, \mn@doi [\apj] {10.3847/1538-4357/aa5ffe}, \href
  {http://adsabs.harvard.edu/abs/2017ApJ...837..170L} {837, 170}

\bibitem[\protect\citeauthoryear{{Madau} \& {Dickinson}}{{Madau} \&
  {Dickinson}}{2014}]{madau2014}
{Madau} P.,  {Dickinson} M.,  2014, \mn@doi [\araa]
  {10.1146/annurev-astro-081811-125615}, \href
  {http://adsabs.harvard.edu/abs/2014ARA%26A..52..415M} {52, 415}

\bibitem[\protect\citeauthoryear{{Mancini}, {Schneider}, {Graziani},
  {Valiante}, {Dayal}, {Maio}  \& {Ciardi}}{{Mancini}
  et~al.}{2016}]{mancini2016}
{Mancini} M.,  {Schneider} R.,  {Graziani} L.,  {Valiante} R.,  {Dayal} P.,
  {Maio} U.,   {Ciardi} B.,  2016, \mn@doi [\mnras] {10.1093/mnras/stw1783},
  \href {http://adsabs.harvard.edu/abs/2016MNRAS.462.3130M} {462, 3130}

\bibitem[\protect\citeauthoryear{{M{\'a}rmol-Queralt{\'o}}, {McLure}, {Cullen},
  {Dunlop}, {Fontana}  \& {McLeod}}{{M{\'a}rmol-Queralt{\'o}}
  et~al.}{2016}]{esther2016}
{M{\'a}rmol-Queralt{\'o}} E.,  {McLure} R.~J.,  {Cullen} F.,  {Dunlop} J.~S.,
  {Fontana} A.,   {McLeod} D.~J.,  2016, \mn@doi [\mnras]
  {10.1093/mnras/stw1212}, \href
  {http://adsabs.harvard.edu/abs/2016MNRAS.460.3587M} {460, 3587}

\bibitem[\protect\citeauthoryear{{McLure} et~al.,}{{McLure}
  et~al.}{2017a}]{mclure2017}
{McLure} R.~J.,  et~al., 2017a, preprint, \href
  {http://adsabs.harvard.edu/abs/2017arXiv170906102M} {} (\mn@eprint {arXiv}
  {1709.06102})

\bibitem[\protect\citeauthoryear{{McLure}, {Pentericci}  \& {VANDELS
  Team}}{{McLure} et~al.}{2017b}]{vandels_messenger}
{McLure} R.,  {Pentericci} L.,   {VANDELS Team} 2017b, The Messenger, \href
  {http://adsabs.harvard.edu/abs/2017Msngr.167...31M} {167, 31}

\bibitem[\protect\citeauthoryear{{Micha{\l}owski}, {Watson}  \&
  {Hjorth}}{{Micha{\l}owski} et~al.}{2010}]{michalowski2010}
{Micha{\l}owski} M.~J.,  {Watson} D.,   {Hjorth} J.,  2010, \mn@doi [\apj]
  {10.1088/0004-637X/712/2/942}, \href
  {http://adsabs.harvard.edu/abs/2010ApJ...712..942M} {712, 942}

\bibitem[\protect\citeauthoryear{{Noll} et~al.,}{{Noll}
  et~al.}{2009}]{noll2009}
{Noll} S.,  et~al., 2009, \mn@doi [\aap] {10.1051/0004-6361/200811526}, \href
  {http://adsabs.harvard.edu/abs/2009A%26A...499...69N} {499, 69}

\bibitem[\protect\citeauthoryear{Oliphant}{Oliphant}{2007}]{oliphant2007}
Oliphant T.~E.,  2007, Computing in Science \& Engineering, 9, 10

\bibitem[\protect\citeauthoryear{{Paardekooper}, {Khochfar}  \& {Dalla
  Vecchia}}{{Paardekooper} et~al.}{2013}]{paardekooper2013}
{Paardekooper} J.-P.,  {Khochfar} S.,   {Dalla Vecchia} C.,  2013, \mn@doi
  [\mnras] {10.1093/mnrasl/sls032}, \href
  {http://adsabs.harvard.edu/abs/2013MNRAS.429L..94P} {429, L94}

\bibitem[\protect\citeauthoryear{{Paardekooper}, {Khochfar}  \& {Dalla
  Vecchia}}{{Paardekooper} et~al.}{2015}]{paardekooper2015}
{Paardekooper} J.-P.,  {Khochfar} S.,   {Dalla Vecchia} C.,  2015, \mn@doi
  [\mnras] {10.1093/mnras/stv1114}, \href
  {http://adsabs.harvard.edu/abs/2015MNRAS.451.2544P} {451, 2544}

\bibitem[\protect\citeauthoryear{Pannella et~al.,}{Pannella
  et~al.}{2015}]{pannella2014}
Pannella M.,  et~al., 2015, \mn@doi [\apj] {10.1088/0004-637X/807/2/141}, \href
  {http://adsabs.harvard.edu/abs/2015ApJ...807..141P} {807, 141}

\bibitem[\protect\citeauthoryear{P\'{e}rez \& Granger}{P\'{e}rez \&
  Granger}{2007}]{perez2007}
P\'{e}rez F.,  Granger B.~E.,  2007, Computing in Science {\&} Engineering, 9,
  21

\bibitem[\protect\citeauthoryear{{Popping}, {Somerville}  \&
  {Galametz}}{{Popping} et~al.}{2017}]{popping2017}
{Popping} G.,  {Somerville} R.~S.,   {Galametz} M.,  2017, \mn@doi [\mnras]
  {10.1093/mnras/stx1545}, \href
  {http://adsabs.harvard.edu/abs/2017MNRAS.471.3152P} {471, 3152}

\bibitem[\protect\citeauthoryear{{Prevot}, {Lequeux}, {Prevot}, {Maurice}  \&
  {Rocca-Volmerange}}{{Prevot} et~al.}{1984}]{prevot1984}
{Prevot} M.~L.,  {Lequeux} J.,  {Prevot} L.,  {Maurice} E.,
  {Rocca-Volmerange} B.,  1984, \aap, \href
  {http://adsabs.harvard.edu/abs/1984A%26A...132..389P} {132, 389}

\bibitem[\protect\citeauthoryear{{Price} et~al.,}{{Price}
  et~al.}{2014}]{price2014}
{Price} S.~H.,  et~al., 2014, \mn@doi [\apj] {10.1088/0004-637X/788/1/86},
  \href {http://adsabs.harvard.edu/abs/2014ApJ...788...86P} {788, 86}

\bibitem[\protect\citeauthoryear{{Reddy} et~al.,}{{Reddy}
  et~al.}{2015}]{reddy2015}
{Reddy} N.~A.,  et~al., 2015, \mn@doi [\apj] {10.1088/0004-637X/806/2/259},
  \href {http://adsabs.harvard.edu/abs/2015ApJ...806..259R} {806, 259}

\bibitem[\protect\citeauthoryear{{Reddy} et~al.,}{{Reddy}
  et~al.}{2017}]{reddy2017}
{Reddy} N.~A.,  et~al., 2017, preprint, \href
  {http://adsabs.harvard.edu/abs/2017arXiv170509302R} {} (\mn@eprint {arXiv}
  {1705.09302})

\bibitem[\protect\citeauthoryear{{Rix}, {Pettini}, {Leitherer}, {Bresolin},
  {Kudritzki}  \& {Steidel}}{{Rix} et~al.}{2004}]{rix2004}
{Rix} S.~A.,  {Pettini} M.,  {Leitherer} C.,  {Bresolin} F.,  {Kudritzki}
  R.-P.,   {Steidel} C.~C.,  2004, \mn@doi [\apj] {10.1086/424031}, \href
  {http://adsabs.harvard.edu/abs/2004ApJ...615...98R} {615, 98}

\bibitem[\protect\citeauthoryear{{Rowlands}, {Gomez}, {Dunne},
  {Arag{\'o}n-Salamanca}, {Dye}, {Maddox}, {da Cunha}  \& {van der
  Werf}}{{Rowlands} et~al.}{2014}]{rowlands2014}
{Rowlands} K.,  {Gomez} H.~L.,  {Dunne} L.,  {Arag{\'o}n-Salamanca} A.,  {Dye}
  S.,  {Maddox} S.,  {da Cunha} E.,   {van der Werf} P.,  2014, \mn@doi
  [\mnras] {10.1093/mnras/stu605}, \href
  {http://adsabs.harvard.edu/abs/2014MNRAS.441.1040R} {441, 1040}

\bibitem[\protect\citeauthoryear{{Salmon} et~al.,}{{Salmon}
  et~al.}{2016}]{salmon2016}
{Salmon} B.,  et~al., 2016, \mn@doi [\apj] {10.3847/0004-637X/827/1/20}, \href
  {http://adsabs.harvard.edu/abs/2016ApJ...827...20S} {827, 20}

\bibitem[\protect\citeauthoryear{{Schaerer}, {Boone}, {Zamojski}, {Staguhn},
  {Dessauges-Zavadsky}, {Finkelstein}  \& {Combes}}{{Schaerer}
  et~al.}{2015}]{schaerer2015}
{Schaerer} D.,  {Boone} F.,  {Zamojski} M.,  {Staguhn} J.,
  {Dessauges-Zavadsky} M.,  {Finkelstein} S.,   {Combes} F.,  2015, \mn@doi
  [\aap] {10.1051/0004-6361/201424649}, \href
  {http://adsabs.harvard.edu/abs/2015A%26A...574A..19S} {574, A19}

\bibitem[\protect\citeauthoryear{{Schreiber} et~al.,}{{Schreiber}
  et~al.}{2015}]{schreiber2015}
{Schreiber} C.,  et~al., 2015, \mn@doi [\aap] {10.1051/0004-6361/201425017},
  \href {http://adsabs.harvard.edu/abs/2015A%26A...575A..74S} {575, A74}

\bibitem[\protect\citeauthoryear{{Scoville}, {Faisst}, {Capak}, {Kakazu}, {Li}
  \& {Steinhardt}}{{Scoville} et~al.}{2015}]{scoville2015}
{Scoville} N.,  {Faisst} A.,  {Capak} P.,  {Kakazu} Y.,  {Li} G.,
  {Steinhardt} C.,  2015, \mn@doi [\apj] {10.1088/0004-637X/800/2/108}, \href
  {http://adsabs.harvard.edu/abs/2015ApJ...800..108S} {800, 108}

\bibitem[\protect\citeauthoryear{{Seon} \& {Draine}}{{Seon} \&
  {Draine}}{2016}]{seon2016}
{Seon} K.-I.,  {Draine} B.~T.,  2016, \mn@doi [\apj]
  {10.3847/1538-4357/833/2/201}, \href
  {http://adsabs.harvard.edu/abs/2016ApJ...833..201S} {833, 201}

\bibitem[\protect\citeauthoryear{{Shapley}, {Steidel}, {Strom},
  {Bogosavljevi{\'c}}, {Reddy}, {Siana}, {Mostardi}  \& {Rudie}}{{Shapley}
  et~al.}{2016}]{shapley2016}
{Shapley} A.~E.,  {Steidel} C.~C.,  {Strom} A.~L.,  {Bogosavljevi{\'c}} M.,
  {Reddy} N.~A.,  {Siana} B.,  {Mostardi} R.~E.,   {Rudie} G.~C.,  2016,
  \mn@doi [\apjl] {10.3847/2041-8205/826/2/L24}, \href
  {http://adsabs.harvard.edu/abs/2016ApJ...826L..24S} {826, L24}

\bibitem[\protect\citeauthoryear{{Speagle}, {Steinhardt}, {Capak}  \&
  {Silverman}}{{Speagle} et~al.}{2014}]{speagle2014}
{Speagle} J.~S.,  {Steinhardt} C.~L.,  {Capak} P.~L.,   {Silverman} J.~D.,
  2014, \mn@doi [\apjs] {10.1088/0067-0049/214/2/15}, \href
  {http://adsabs.harvard.edu/abs/2014ApJS..214...15S} {214, 15}

\bibitem[\protect\citeauthoryear{{Stanway}, {Eldridge}  \& {Becker}}{{Stanway}
  et~al.}{2016}]{stanway2016}
{Stanway} E.~R.,  {Eldridge} J.~J.,   {Becker} G.~D.,  2016, \mn@doi [\mnras]
  {10.1093/mnras/stv2661}, \href
  {http://adsabs.harvard.edu/abs/2016MNRAS.456..485S} {456, 485}

\bibitem[\protect\citeauthoryear{{Stecher}}{{Stecher}}{1965}]{stecher1965}
{Stecher} T.~P.,  1965, \mn@doi [\apj] {10.1086/148462}, \href
  {http://adsabs.harvard.edu/abs/1965ApJ...142.1683S} {142, 1683}

\bibitem[\protect\citeauthoryear{{Steidel}, {Strom}, {Pettini}, {Rudie},
  {Reddy}  \& {Trainor}}{{Steidel} et~al.}{2016}]{steidel2016}
{Steidel} C.~C.,  {Strom} A.~L.,  {Pettini} M.,  {Rudie} G.~C.,  {Reddy} N.~A.,
    {Trainor} R.~F.,  2016, \mn@doi [\apj] {10.3847/0004-637X/826/2/159}, \href
  {http://adsabs.harvard.edu/abs/2016ApJ...826..159S} {826, 159}

\bibitem[\protect\citeauthoryear{{Strom}, {Steidel}, {Rudie}, {Trainor},
  {Pettini}  \& {Reddy}}{{Strom} et~al.}{2017}]{strom2017}
{Strom} A.~L.,  {Steidel} C.~C.,  {Rudie} G.~C.,  {Trainor} R.~F.,  {Pettini}
  M.,   {Reddy} N.~A.,  2017, \mn@doi [\apj] {10.3847/1538-4357/836/2/164},
  \href {http://adsabs.harvard.edu/abs/2017ApJ...836..164S} {836, 164}

\bibitem[\protect\citeauthoryear{{Vanzella} et~al.,}{{Vanzella}
  et~al.}{2016}]{vanzella2016}
{Vanzella} E.,  et~al., 2016, \mn@doi [\apj] {10.3847/0004-637X/825/1/41},
  \href {http://adsabs.harvard.edu/abs/2016ApJ...825...41V} {825, 41}

\bibitem[\protect\citeauthoryear{{Watson}, {Christensen}, {Knudsen}, {Richard},
  {Gallazzi}  \& {Micha{\l}owski}}{{Watson} et~al.}{2015}]{watson2015}
{Watson} D.,  {Christensen} L.,  {Knudsen} K.~K.,  {Richard} J.,  {Gallazzi}
  A.,   {Micha{\l}owski} M.~J.,  2015, \mn@doi [\nat] {10.1038/nature14164},
  \href {http://adsabs.harvard.edu/abs/2015Natur.519..327W} {519, 327}

\bibitem[\protect\citeauthoryear{{Wild}, {Charlot}, {Brinchmann}, {Heckman},
  {Vince}, {Pacifici}  \& {Chevallard}}{{Wild} et~al.}{2011}]{wild2011}
{Wild} V.,  {Charlot} S.,  {Brinchmann} J.,  {Heckman} T.,  {Vince} O.,
  {Pacifici} C.,   {Chevallard} J.,  2011, \mn@doi [\mnras]
  {10.1111/j.1365-2966.2011.19367.x}, \href
  {http://adsabs.harvard.edu/abs/2011MNRAS.417.1760W} {417, 1760}

\bibitem[\protect\citeauthoryear{{Zafar}, {Watson}, {Fynbo}, {Malesani},
  {Jakobsson}  \& {de Ugarte Postigo}}{{Zafar} et~al.}{2011}]{zafar2011}
{Zafar} T.,  {Watson} D.,  {Fynbo} J.~P.~U.,  {Malesani} D.,  {Jakobsson} P.,
  {de Ugarte Postigo} A.,  2011, \mn@doi [\aap] {10.1051/0004-6361/201116663},
  \href {http://adsabs.harvard.edu/abs/2011A%26A...532A.143Z} {532, A143}

\bibitem[\protect\citeauthoryear{{Zeimann} et~al.,}{{Zeimann}
  et~al.}{2015}]{zeimann2015_dust}
{Zeimann} G.~R.,  et~al., 2015, \mn@doi [\apj] {10.1088/0004-637X/814/2/162},
  \href {http://adsabs.harvard.edu/abs/2015ApJ...814..162Z} {814, 162}

\bibitem[\protect\citeauthoryear{{de Barros} et~al.,}{{de Barros}
  et~al.}{2016}]{debarros2016}
{de Barros} S.,  et~al., 2016, \mn@doi [\aap] {10.1051/0004-6361/201527046},
  \href {http://adsabs.harvard.edu/abs/2016A%26A...585A..51D} {585, A51}

\makeatother
\end{thebibliography}


\label{lastpage}
\bsp
\end{document}